\pdfoutput=1
\documentclass[12pt,letterpaper]{article}
\usepackage[a4paper, total={7in, 10in}]{geometry}

\usepackage{graphicx}
\usepackage{float}
\usepackage{helvet}
\usepackage{authblk}
\usepackage{hyperref}
\usepackage{amsmath}
\usepackage{amssymb}
\usepackage{xcolor}
\usepackage{orcidlink}
\usepackage[super,comma,sort&compress]
   {natbib}
\usepackage{bibunits}
\defaultbibliography{references}
\defaultbibliographystyle{numbered}

\usepackage{eurosym}
\usepackage{subcaption}
\usepackage{parskip}
\usepackage{setspace}
\newcounter{sisec}
\renewcommand{\thesisec}{S\arabic{sisec}}
\newcommand{\sisection}[1]{\refstepcounter{sisec}\subsection*{\thesisec. #1}}
\newcommand{\anchor}[1]{\phantomsection\label{#1}}

\makeatletter
\renewcommand{\maketitle}{\bgroup\setlength{\parindent}{0pt}
\begin{flushleft}
  \textbf{\@title}

  \@author
\end{flushleft}\egroup}
\makeatother

\title{Time to AC-t: When AC-transmission changes capacity expansion planning}
\date{}

\author[1,4,*]{Lara Booth}
\author[2]{Shiva Madadkhani}
\author[3,*]{Ruaridh Macdonald}
\author[1, *]{Priya L. Donti}

\affil[1]{Department of Electrical Engineering \& Computer Science and Laboratory for Information \& Decision Systems, Massachusetts Institute of Technology, Cambridge, USA}
\affil[2]{Center for Energy Markets, School of Management, Technical University of Munich, Germany}
\affil[3]{MIT Energy Initiative, Massachusetts Institute of Technology, Cambridge, USA}

\affil[4]{Lead contact}
\affil[*]{Correspondence: lcbooth@mit.edu, rmacd@mit.edu, donti@mit.edu}

\begin{document}
\begin{bibunit}

\maketitle

\section*{SUMMARY}

Capacity expansion models often dramatically simplify the physics of how power flows through a transmission network in order to preserve computational tractability. However, with the rapid growth of electrification, large loads, renewables, and storage, increased realism in planning future grids is imperative. We extend the open-source capacity expansion model GenX with AC optimal power flow physics, and analyze generation, storage, and transmission investment decisions for 20-zone, hourly models of Germany and Texas. We compare our results against those using two common approximations, the transport model and DC optimal power flow. We find that total costs and emissions largely agree across all formulations. However, we find substantial differences in outputs related to transmission, generation siting, storage, and reactive-power support between the approximate formulations and the true AC physics formulation. We identify where AC physics changes planning decisions and discuss how to potentially capture these effects without the full computational cost.

\section*{CONTEXT \& SCALE}

Electricity grids are changing rapidly. Electrification is increasing demand, new large loads such as data centers are connecting to the grid, and variable renewable resources such as wind and solar, paired with battery storage, make up a growing share of supply. Adapting to these changes will require major investments in new generation, storage, and transmission lines over the coming decades, and the transmission network is already a bottleneck in many regions. 

To plan these investments, grid planners use tools called capacity expansion models to decide what to build and where. 
To keep computation times manageable, these models usually adopt simplified representations of how
power flows through the network. Many models route power along any transmission line that has sufficient capacity, when in reality power spreads across every available path according to the network's electrical properties. Such simplifications can produce investment plans that are not fully compatible with true power grid physics, requiring additional remedial investments to enable grids to operate reliably under real-world power flow.

Our work adds full alternating-current power flow physics to an open-source capacity expansion model, and compares the results for case studies of Germany and Texas against those under two common simplified transmission representations. We find that system-wide costs and emissions agree across all three transmission representations, but that this does not tell the whole story. Importantly, the models differ in where new generation and transmission get built and in how much battery storage is needed. These differences appear largest where generation is far from demand, an increasingly common situation as renewables' share of supply expands. Because planning models with full power flow remain slow to solve, we discuss how adding targeted constraints to the faster simplified models could capture relevant effects without adding undue computational cost.

\section*{KEYWORDS}
capacity expansion model, transmission expansion, acopf, dcopf, power systems planning

\newpage
\section*{INTRODUCTION}
Electricity systems worldwide are undergoing rapid transformation. On the supply side, the falling cost of low-emission generation and storage\cite{way2022empirically} along with fuel price volatility is changing the generation mix. On the demand side, rising electricity consumption in emerging markets,\cite{iea2026electricity} increasing electrification of end-use sectors,\cite{luderer2022impact, mallapragada2023decarbonization} and the emergence of new large loads such as data centers\cite{nerc_large_loads_2025} are reshaping demand profiles and peak-load patterns. The transmission and distribution networks that connect supply and demand are also challenged by rising congestion and investment barriers.\cite{gorman2025grid, rand2026queued} Against this backdrop, balancing supply and demand reliably and affordably requires careful long-term system planning alongside regulatory and policy frameworks to align market incentives with broader system-wide objectives. 

Capacity expansion models (CEMs) are a central tool for long-term planning, identifying the least-cost combination of generation, storage, and (in some cases) transmission investments needed to meet projected demand.\cite{schivley2025process, schlachtberger2017benefits, schlachtberger2018cost, neumann2021near, brown2021value, pineda2018chronological} Like any model, CEMs must balance fidelity with computational tractability. While spatial and temporal resolution are classic dimensions of this trade-off,\cite{horsch2017role, schyska2021sensitivity, pecora2025quantifying} the representation of the power transmission network also affects both model results and runtime.\cite{brown2021value} At the high-fidelity end is AC optimal power flow (ACOPF). This representation captures all steady-state network physics, including how reactive power must be supplied to keep voltages within their limits. The resulting constraints are nonlinear and nonconvex, making any CEM that incorporates ACOPF rather expensive to solve in long-term planning contexts.\cite{bienstock2019strong, weng2026modeling} CEMs therefore use simplified transmission representations that reduce the computational burden. These include DC optimal power flow (DCOPF),\cite{neumann2022assessments, LeeSun2025_GTEP} which neglects reactive power and voltage magnitudes, and the transport model,\cite{jenkins2017enhanced} which only limits power flows by per-line capacities. Today, the transport model is the default in many open-source planning tools.\cite{genx, REEDS, brown2017pypsa, macdonald2025macroenergy}

Although simplifying the network representation improves tractability, an important question remains unresolved: to what extent does it distort planning outcomes? Previous studies in both generation expansion and, separately, transmission expansion have found varying results. On the generation side, \citet{weng2026modeling} compare the transport model against ACOPF, finding significant differences in generation capacity investments for a 3-zone model of ISO New England. \citet{wogrin2020assessing} add reactive power constraints to generation expansion planning through a second-order cone approximation of AC power flow; they show that omitting these detailed transmission constraints can distort optimal resource allocation and cause infeasibilities. On the transmission side, \citet{bent2014bridging} find that DC transmission expansion plans substantially underestimate expansion costs and violate AC line and voltage limits when evaluated on the full network. \citet{zhang2013improved}, who augment a transmission expansion model with a linearized representation of reactive power and losses, similarly find that DC-based plans can require additional network reinforcements. In their model of Egypt's transmission network, \citet{farrag2019ac} find that AC selects fewer lines than DC formulations. In contrast, \citet{khanpour2026empirical} report close agreement between AC- and DC-based transmission investment costs. These generation and transmission expansion studies provide important insights, but most focus on small test networks, and none model renewables-rich systems where generation, transmission, and storage all compete for investment. 

In this work, we ask how the representation of network physics shapes planning outcomes in larger systems with substantial renewables and storage, looking beyond only aggregate costs to also the specific assets that are built. We systematically compare three power flow formulations of increasing physical realism -- the transport model, DCOPF, and full ACOPF -- across two 20-zone systems representing Germany and the Electric Reliability Council of Texas (ERCOT). 

To conduct this study, we extend the open-source CEM GenX\cite{genx} with full AC power flow constraints. Our experiments co-optimize generation, storage, and transmission investment under AC power flow across 20 zones and a year of hourly operation. In our framework, we develop a representation of reactive power for CEMs, deriving each resource's reactive capability from grid-code power factor requirements. We also extend GenX's DCOPF formulation to allow for transmission expansion, updating impedances as line capacity is added, consistent with ACOPF. Finally, while neither the transport model nor DCOPF inherently represents losses, common practice is to approximate them linearly as a fixed fraction of flow;\cite{genx, brown2017pypsa} we adopt this approach for both approximate formulations to allow a more fair comparison with ACOPF.

We find that the three formulations agree on total system cost to within 1.2\%, but differ considerably in where generation is built and how much storage and transmission are installed. In Germany, total generation capacity differs from ACOPF by only 2.1\% under transport and 1.2\% under DC, yet the median zonal capacity differences are 37.5\% and 12.0\%, respectively. Some of these differences arise because the transport model does not capture competition between renewable generators for shared transmission capacity, allowing more investment at sites whose output would otherwise contribute to congestion. Reactive power requirements also affect investment decisions. Under our baseline assumptions, ACOPF builds more battery power capacity, with shorter storage duration, to supply reactive power alongside batteries' use for energy arbitrage. The size of this effect depends on how we model reactive capability and which technologies can provide it. These results suggest that simplified transmission models can give similar system-wide results while selecting quite different investments, an important distinction when using CEMs to inform generation siting and network expansion.

\section*{RESULTS}
\textbf{Experimental setup}

In our analysis, we use the open-source model GenX\cite{genx} that co-optimizes capacity deployment and retirement decisions with hourly generation dispatch to minimize total costs over a given planning horizon. Using GenX, we run a single-year model that co-optimizes generation, storage, and transmission expansion to compare three transmission representations of increasing physical detail: the transport model, DCOPF, and ACOPF.

We construct input datasets for 20-zone representations of Germany and ERCOT using PyPSA-Eur\cite{PyPSAEur} and PyPSA-USA,\cite{Tehranchi_PyPSA-USA_An_Open-Source} respectively. These datasets include detailed transmission line parameters, hourly demand profiles, hourly renewable capacity factors, and zonal renewable capacity potentials, which we convert into a GenX-compatible format (see \hyperref[methods:data]{Methods, \emph{Data collection and pre-processing}}). We group individual generators into technology classes. We consider onshore wind, utility-scale solar photovoltaic, open-cycle gas turbines (OCGT), combined-cycle gas turbines (CCGT), and battery storage as technology options in both systems, and additionally we include biomass and offshore wind in Germany and coal generation in ERCOT. The asymmetry in generation options is because offshore wind and biomass are not significant sources of generation in ERCOT, and because Germany plans to phase out coal in the near future\cite{kvbg2020} (see \hyperref[methods:data]{Methods, \emph{Data collection and pre-processing}}). We note that direct comparisons of investment levels across Germany and ERCOT should be interpreted with caution, because technology cost assumptions differ between the U.S. and European input datasets (see Table~\ref{tabS:inputs}). 

To reduce the runtime of GenX with the ACOPF transmission representation, we use time domain reduction (TDR) to cluster the full annual time series into a smaller set of representative periods whose weighted sum reproduces a full year of operation. In \ref{si:tdr}, we test different numbers of representative periods by comparing their deviation from the full-year transport solution against the associated solve times, and select the smallest count that reproduces full-year system cost, emissions, and installed capacity by resource type to within 2\% error. This yielded 32 periods for ERCOT and 20 for Germany (see Fig.~\ref{fig:TDR}). We spatially aggregate each network, with thousands of buses, into 20 zones, and calculate aggregate existing transmission capacity between each zone. 
We model generation and storage investment using a greenfield approach (i.e., assuming no existing generation or storage capacity), 
to facilitate direct comparisons of investment between transmission formulations. Our systems do include an existing transmission network that can be reinforced jointly with generation expansion.  

\textbf{Spatiotemporal differences in supply and demand across modeled systems.}

\begin{figure}[htbp]
    \centering
    \includegraphics[width=\linewidth]{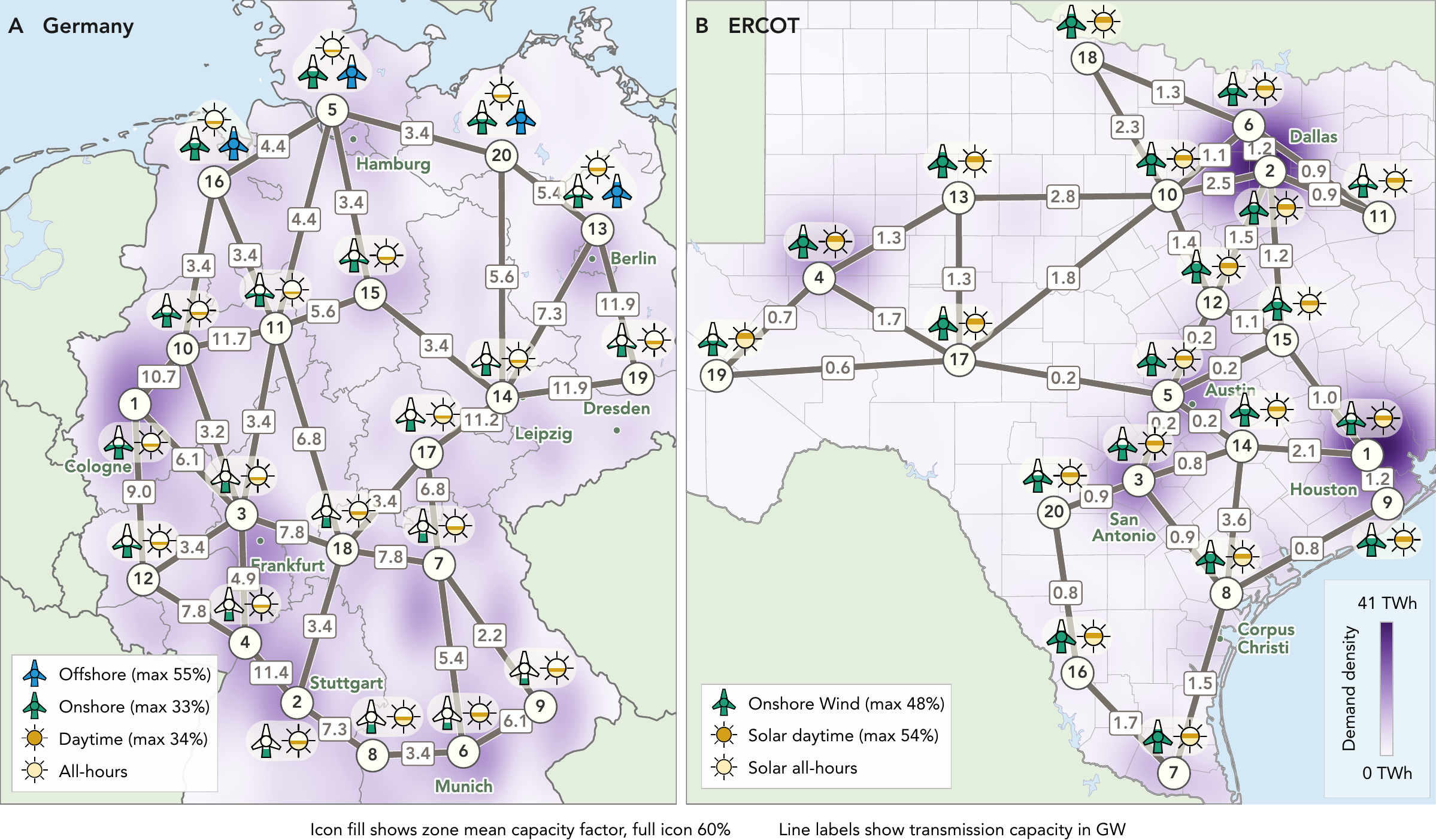}
    \caption{\textbf{Zonal VRE resource availability and transmission network.}
    (A) German 20-zone system. (B) ERCOT 20-zone system. Icon fill indicates each zone's time-weighted mean capacity factor for that resource. For both systems, a full icon represents 60\%, since no zones were observed to have higher than 55\% mean capacity factor. Solar fill is two-tone: pale yellow indicates the all-hours mean, and pale yellow plus darker gold indicate the daytime-only mean, on the same 60\% scale. Both plots share one demand-density color scale.}
    \label{fig:vre_maps}
\end{figure}

Our results depend on the spatial and temporal distribution of demand and resources, and thus we begin by outlining their differences in our two test systems. We select Germany and ERCOT as two regions undergoing substantial renewables expansion but differing in renewable resources, demand patterns, and grid topology. As seen in Fig. \ref{fig:vre_maps}, Germany, which is roughly half the geographic area of Texas, has a more evenly distributed electricity demand in our test dataset. In the ERCOT dataset, by contrast, demand is concentrated around major urban areas in the east and along the coast, with additional industrial demand in the Permian Basin in the west. This pattern reflects differences in the spatial concentration of population and industrial activity between the two regions, with large parts of West Texas being sparsely populated.\cite{swanson2025new} The two largest ERCOT zones at Houston and Dallas carry 43\% of annual demand (compared with 17\% for Germany's two largest zones). This concentration, together with low gas prices, also shapes where new generation is built. Without a limit on thermal additions, our capacity expansion model builds roughly 14\,GW of new gas capacity in each of Houston and Dallas, which is inconsistent with ERCOT's siting practice and current development pipeline (see \ref{si:caps}). We therefore cap new thermal capacity at 5\,GW per zone in all ERCOT results. Beyond spatial differences, seasonal demand patterns also differ between the two cases. Electricity demand peaks during the cooler winter months in Germany, whereas in Texas, it is driven more strongly by cooling needs and peaks during the warmer summer months (see Fig.~\ref{figS:demand}). 

Renewable resource availability also differs markedly between the regions. Germany shows a clear north--south difference in wind availability, with the highest wind potential in the north, particularly offshore wind, located far from major load centers in the west and south. By contrast, resources supplying generation to ERCOT's load centers at Dallas and Houston are more evenly distributed within the system. Wind capacity-factor time series are more highly correlated across German regions than across ERCOT (see Fig.~\ref{figS:wind_corr}). Yet, wind generation shows greater regional complementarity in ERCOT, for example between southern and western Texas.\cite{slusarewicz2018assessing} For our selected weather years (2023 for Germany and 2022 for Texas), the mean onshore wind capacity factor across zones is 25\% in Germany against 38\% in Texas, and the mean daytime solar capacity factor is 31\% against 45\%. Spatial variation within each system is much larger for wind than for solar: zonal onshore wind capacity factors range from 16--34\% in Germany and 27--48\% in Texas, compared with 30--33\% and 42--54\% for daytime solar. German offshore wind, on the other hand, has capacity factors that exceed those of the best Texan renewable resource. Although our main results compare transmission formulations within each system rather than between systems, these regional differences shape where and how transmission representation matters.

\textbf{System-wide costs and emissions agree closely across transmission formulations.}

Table~\ref{tab:headline} shows the system-wide costs of meeting electricity demand in each jurisdiction, corresponding to the objective function of the CEM optimization. These costs comprise generation and transmission investment costs, operational costs across all time steps, and the cost of non-served energy. Operational costs are evaluated using each model’s transmission approximation; the resulting investment decisions are not subsequently evaluated using AC power flow. Across the three transmission formulations, total costs remain within 1.2\% of one another for both the Germany and ERCOT systems. Total system-wide CO$_2$ emissions are likewise very similar, although they are slightly higher under the DC formulation. Across the three transmission formulations, the broad spatial pattern of capacity buildouts follows resource availability: wind capacity is concentrated in northern Germany and the wind-rich regions of Texas, while solar PV is distributed across more zones in both systems. 

\begin{table}[htbp]
\centering
\caption{\textbf{System cost and capacity comparison: Transport and DCOPF approximations vs.\ ACOPF.} Network expansion; lossy; 20-period TDR for Germany, 32-period TDR for ERCOT. Demand is an input and identical across transmission network formulations. Storage discharge is not netted with charging, so total generation exceeds demand by storage charging and transmission losses. Deviations from ACOPF of 5\% or more are shown in bold.}
\label{tab:headline}

\begin{minipage}[t]{0.48\linewidth}
\centering
\scriptsize
\setlength{\tabcolsep}{3pt}
\setlength{\parskip}{0pt}
\textbf{Germany}

\medskip
\begin{tabular}{|l | r | r | r|}
\hline
\textbf{Cost (billion \EUR{})} & \textbf{Transport} & \textbf{DCOPF} & \textbf{ACOPF} \\ [1ex]
\hline
Fixed & 25.05 & 25.10 & 25.30 \\ [1ex]
\hline
Fuel & 12.21 & 12.64 & 12.05 \\ [1ex]
\hline
Variable O\&M & 1.08 & 1.10 & 1.05 \\ [1ex]
\hline
Network expansion & 0.23 & 0.24 & 0.26 \\ [1ex]
\hline
Non-served energy & 0.10 & 0.04 & 0.00 \\ [1ex]
\hline
\textbf{Total} & 38.67 & 39.13 & 38.66 \\ [1ex]
\hline
Emissions (Mt CO$_2$) & 60.3 & 62.5 & 59.6 \\ [1ex]
\hline\hline
\multicolumn{4}{|l|}{\textbf{Capacity (GW)}} \\ [1ex]
\hline
Thermal (CCGT + OCGT) & 64.7 & 64.3 & 62.6 \\
Wind (onshore + offshore) & 72.5 & 70.0 & 69.3 \\
Solar & 97.8 & 101.9 & 99.0 \\
Storage (battery) & 7.3 & 8.1 & 16.6 \\
\quad\textit{energy, GWh} & 21.4 & 22.9 & 26.5 \\
\quad\textit{duration, h} & 2.91 & 2.82 & 1.60 \\
\hline
Total capacity & 242.3 & 244.3 & 247.4 \\ [1ex]
\hline
New transmission capacity, GVA & 30.3 & 31.2 & 34.2 \\ [1ex]
\hline\hline
\multicolumn{4}{|l|}{\textbf{Generation (TWh)}} \\ [1ex]
\hline
Demand & \multicolumn{3}{c|}{505.4} \\
\hline
Thermal (CCGT + OCGT) & 161.3 & 167.0 & 159.8 \\
Wind (onshore + offshore) & 234.7 & 225.9 & 229.0 \\
Solar & 127.2 & 131.4 & 128.9 \\
Storage discharge & 7.1 & 9.0 & 9.2 \\
\hline
Total generation & 530.4 & 533.3 & 526.9 \\ [1ex]
\hline\hline
\multicolumn{4}{|l|}{\textbf{\% error vs.\ AC reference}} \\ [1ex]
\hline
Total cost & +0.03\% & +1.20\% & -- \\ [1ex]
\hline
Emissions & +1.3\% & +4.9\% & -- \\ [1ex]
\hline
Thermal, capacity & +3.4\% & +2.8\% & -- \\
\quad\textit{generation} & +0.9\% & +4.5\% & -- \\ [1ex]
\hline
Wind, capacity & +4.6\% & +1.0\% & -- \\
\quad\textit{generation} & +2.5\% & -1.4\% & -- \\ [1ex]
\hline
Solar, capacity & -1.2\% & +2.9\% & -- \\
\quad\textit{generation} & -1.3\% & +2.0\% & -- \\ [1ex]
\hline
Storage, capacity & \textbf{-55.7\%} & \textbf{-50.9\%} & -- \\
\quad\textit{discharge} & \textbf{-22.2\%} & -1.9\% & -- \\ [1ex]
\hline
New transmission capacity & \textbf{-11.3\%} & \textbf{-8.7\%} & -- \\ [1ex]
\hline
\end{tabular}
\end{minipage}
\hfill
\begin{minipage}[t]{0.48\linewidth}
\centering
\scriptsize
\setlength{\tabcolsep}{3pt}
\setlength{\parskip}{0pt}
\textbf{ERCOT}

\medskip
\begin{tabular}{|l | r | r | r|}
\hline
\textbf{Cost (billion \$)} & \textbf{Transport} & \textbf{DCOPF} & \textbf{ACOPF} \\ [1ex]
\hline
Fixed & 11.08 & 11.04 & 11.21 \\ [1ex]
\hline
Fuel & 7.49 & 7.58 & 7.40 \\ [1ex]
\hline
Variable O\&M & 0.44 & 0.44 & 0.43 \\ [1ex]
\hline
Network expansion & 0.11 & 0.12 & 0.11 \\ [1ex]
\hline
Non-served energy & 0.14 & 0.14 & 0.12 \\ [1ex]
\hline
\textbf{Total} & 19.25 & 19.32 & 19.27 \\ [1ex]
\hline
Emissions (Mt CO$_2$) & 109.2 & 110.5 & 107.9 \\ [1ex]
\hline\hline
\multicolumn{4}{|l|}{\textbf{Capacity (GW)}} \\ [1ex]
\hline
Thermal (CCGT + OCGT) & 58.5 & 59.0 & 58.4 \\
Wind (onshore) & 22.3 & 21.2 & 22.8 \\
Solar & 58.2 & 59.2 & 59.0 \\
Storage (battery) & 1.6 & 1.6 & 2.6 \\
\quad\textit{energy, GWh} & 2.5 & 2.5 & 3.2 \\
\quad\textit{duration, h} & 1.59 & 1.61 & 1.25 \\
\hline
Total capacity & 140.6 & 141.0 & 142.7 \\ [1ex]
\hline
New transmission capacity, GVA & 11.3 & 11.4 & 10.9 \\ [1ex]
\hline\hline
\multicolumn{4}{|l|}{\textbf{Generation (TWh)}} \\ [1ex]
\hline
Demand & \multicolumn{3}{c|}{429.9} \\
\hline
Thermal (CCGT + OCGT) & 237.8 & 240.7 & 235.0 \\
Wind (onshore) & 81.7 & 77.0 & 81.6 \\
Solar & 113.7 & 115.5 & 115.4 \\
Storage discharge & 0.5 & 0.6 & 0.7 \\
\hline
Total generation & 433.8 & 433.8 & 432.7 \\ [1ex]
\hline\hline
\multicolumn{4}{|l|}{\textbf{\% error vs.\ AC reference}} \\ [1ex]
\hline
Total cost & -0.09\% & +0.25\% & -- \\ [1ex]
\hline
Emissions & +1.2\% & +2.4\% & -- \\ [1ex]
\hline
Thermal, capacity & +0.2\% & +1.0\% & -- \\
\quad\textit{generation} & +1.2\% & +2.4\% & -- \\ [1ex]
\hline
Wind, capacity & -1.9\% & \textbf{-6.7\%} & -- \\
\quad\textit{generation} & +0.1\% & \textbf{-5.6\%} & -- \\ [1ex]
\hline
Solar, capacity & -1.4\% & +0.3\% & -- \\
\quad\textit{generation} & -1.4\% & +0.1\% & -- \\ [1ex]
\hline
Storage, capacity & \textbf{-37.6\%} & \textbf{-38.4\%} & -- \\
\quad\textit{discharge} & \textbf{-26.8\%} & \textbf{-17.3\%} & -- \\ [1ex]
\hline
New transmission capacity & +3.6\% & +4.6\% & -- \\ [1ex]
\hline
\end{tabular}
\end{minipage}
\end{table}

\textbf{DCOPF and transport models site capacity differently.}

\begin{figure}[p]
    \centering
    \includegraphics[width=\linewidth]{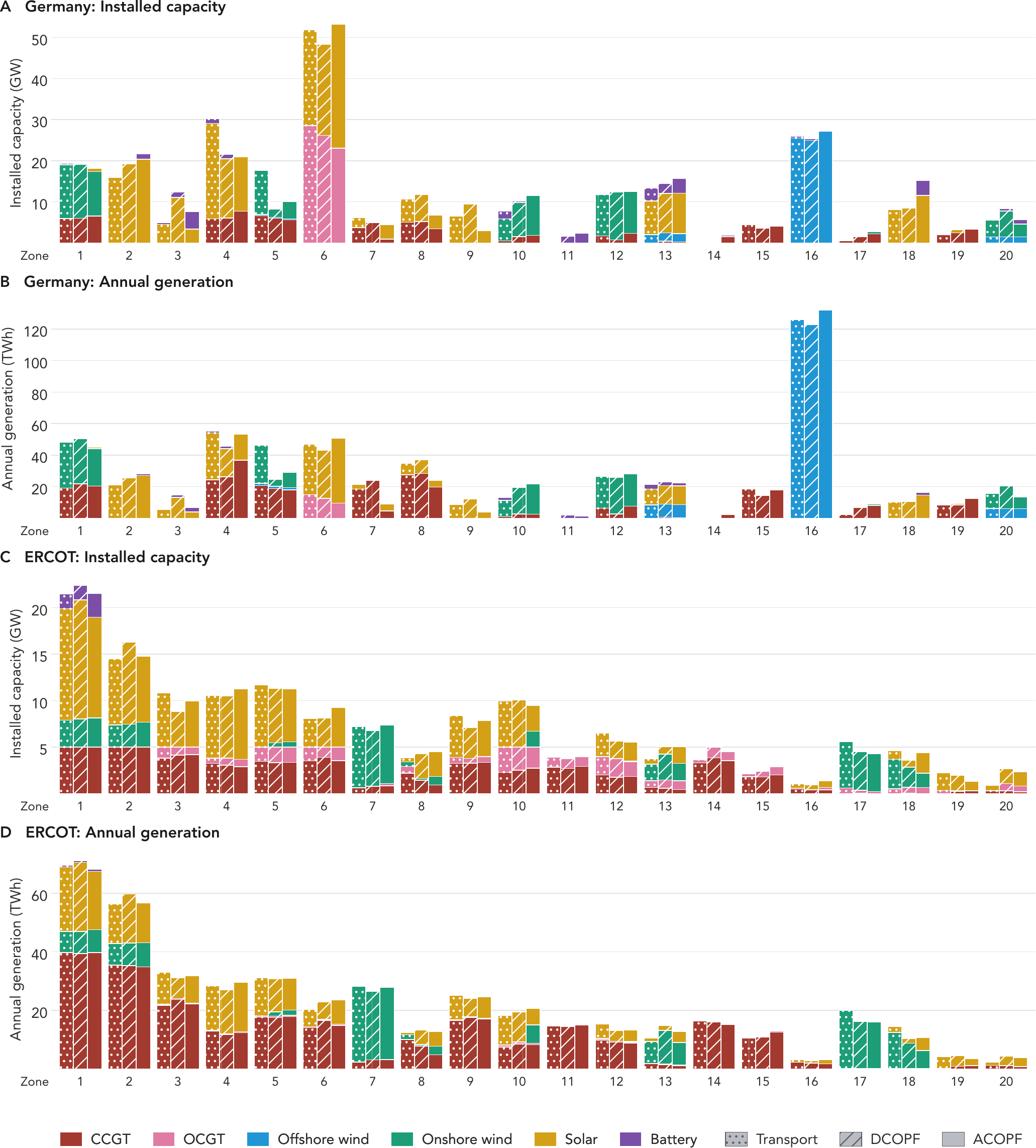}
    \caption{\textbf{Zonal capacity and generation by transmission model.}
    (A, B) Germany; (C, D) ERCOT under the 5\,GW per-zone thermal cap. Bars are
    stacked by technology and grouped by formulation within each zone: ACOPF
    solid, DCOPF diagonal hatch, Transport dotted. Battery generation is gross
    discharge.}
    \label{fig:zone_bars}
\end{figure}

While the transmission formulations agree on system-level costs and emissions, the models diverge in where generation capacity is sited, which lines are expanded, and how much battery storage is built. Fig.~\ref{fig:zone_bars} breaks down capacity investments and annual generation by technology and zone. For both Germany and ERCOT, the set of technologies selected in each zone is largely unchanged across model formulations. What varies instead is how much of each technology is built per zone, and these differences are unevenly distributed, with no formulation being consistently higher or lower than the others across zones (Table~\ref{tab:zonal_stats}). In Germany, the largest deviations in zonal generation capacity from ACOPF are $-$6.6 GW in zone 18 compared to DCOPF and $+$9.3 GW in zone 4 compared to transport. Two zones (11 and 14) that are allocated generation capacity under ACOPF receive no generation capacity under transport (Fig.~\ref{fig:zone_bars}A). ERCOT shows the same behavior at a smaller absolute scale, with its largest capacity differences from ACOPF being $+$1.5 GW in zone 2 under DCOPF and $-$1.5 GW in zone 20 under transport (Fig.~\ref{fig:zone_bars}C). Relative deviations are also larger in Germany: median zonal capacity differs from ACOPF by 37.5\% under transport and 12.0\% under DCOPF, compared with 10.8\% and 8.7\% in ERCOT. 

Close agreement in total capacity hides large differences in where that capacity is built. In Germany, total installed capacity under the transport and DCOPF formulations is within 2.1 and 1.2\% of ACOPF, yet the absolute differences from ACOPF across all zone and technology pairs sum to 35.9 and 31.8\% of ACOPF's total installed capacity, respectively. In ERCOT, totals are within 1.5 and 1.2\%, while the zone and technology differences sum to 22.6 and 17.1\%. Surpluses in some zones offset shortfalls in others, so the approximate transmission formulations appear accurate at the system level while producing substantial differences in how capacity gets sited.

Annual generation follows the same pattern (Fig.~\ref{fig:zone_bars}B,~D). In Germany, median zonal generation differs from ACOPF by 29.1\% under transport and 22.6\% under DCOPF, and individual zones differ by far more: zone 7 generates 166\% more under DCOPF than under ACOPF. In ERCOT, the median zonal differences are 6.0\% and 4.8\%, but zone 20 generates 41\% less under transport. Meanwhile, the total annual generation agrees to within about 1\% across formulations in both systems (Table~\ref{tab:headline}). These zonal differences in generation are shaped by how each formulation represents power flow through the transmission network, which we examine next.

\begin{table}[htbp]
\centering
\caption{\textbf{Zonal and per-line disagreement with ACOPF.} For each formulation, the median absolute difference from ACOPF and the largest difference across zones (or, in the last block, across transmission lines). Worst cases are positive where the approximation exceeds ACOPF, negative where it falls short. The third column gives the zone, or for lines the two zones the line connects (e.g.\ 5--20), where the worst case occurs. ERCOT wind is onshore only. Zone numbers are as in Fig.~\ref{fig:vre_maps}. \emph{Total, \%} rows give the difference in a zone's total capacity or generation as a percentage of its ACOPF value. \emph{Hours $>$80\%} is a measure of congestion given by the difference in the percentage of hours a line is loaded above 80\% of its rating, in percentage points (pp).}
\label{tab:zonal_stats}

\begin{minipage}[t]{0.495\linewidth}
\centering
\scriptsize
\setlength{\tabcolsep}{1.9pt}
\setlength{\parskip}{0pt}
\textbf{\strut Germany}

\medskip
\begin{tabular}{|l | c c c | c c c|}
\hline
 & \multicolumn{3}{c|}{\textbf{Transport}} & \multicolumn{3}{c|}{\textbf{DCOPF}} \\
\textbf{vs.\ ACOPF} & median & worst & zone & median & worst & zone \\[-0.8ex]
 & {\tiny (absolute)} & {\tiny (signed)} & {\tiny or line} & {\tiny (absolute)} & {\tiny (signed)} & {\tiny or line} \\
\hline
\multicolumn{7}{|l|}{\textbf{Zonal capacity}} \\
\hline
Total, \% & 37.5 & +118.4 & 9 & 12.0 & +212.9 & 9 \\
Thermal, GW & 1.37 & +5.50 & 6 & 0.94 & +4.00 & 7 \\
Wind, GW & 1.22 & +6.54 & 5 & 1.81 & +3.17 & 20 \\
Solar, GW & 2.92 & +9.97 & 4 & 3.06 & $-$7.94 & 6 \\
Storage, GW & 1.15 & $-$3.91 & 3 & 0.68 & $-$3.56 & 18 \\
\hline
\multicolumn{7}{|l|}{\textbf{Zonal generation}} \\
\hline
Total, \% & 29.1 & +137.1 & 7 & 22.6 & +213.7 & 9 \\
Thermal, TWh & 2.60 & +13.75 & 7 & 3.11 & +19.31 & 7 \\
Wind, TWh & 3.62 & +14.45 & 5 & 3.72 & $-$9.18 & 16 \\
Solar, TWh & 3.62 & +13.14 & 4 & 3.87 & $-$10.82 & 6 \\
Storage, TWh & 1.22 & $-$2.35 & 3 & 0.54 & $-$1.61 & 18 \\
\hline
\multicolumn{7}{|l|}{\textbf{Per line}} \\
\hline
New capacity, GVA & 0.53 & $-$2.87 & 10–3 & 0.31 & $-$3.40 & 5–20 \\
Hours $>$80\%, pp & 7.1 & +44.8 & 18–11 & 0.3 & +7.1 & 5–20 \\
\hline
\end{tabular}
\end{minipage}
\hfill
\begin{minipage}[t]{0.495\linewidth}
\centering
\scriptsize
\setlength{\tabcolsep}{1.9pt}
\setlength{\parskip}{0pt}
\textbf{\strut ERCOT}

\medskip
\begin{tabular}{|l | c c c | c c c|}
\hline
 & \multicolumn{3}{c|}{\textbf{Transport}} & \multicolumn{3}{c|}{\textbf{DCOPF}} \\
\textbf{vs.\ ACOPF} & median & worst & zone & median & worst & zone \\[-0.8ex]
 & {\tiny (absolute)} & {\tiny (signed)} & {\tiny or line} & {\tiny (absolute)} & {\tiny (signed)} & {\tiny or line} \\
\hline
\multicolumn{7}{|l|}{\textbf{Zonal capacity}} \\
\hline
Total, \% & 10.8 & +69.9 & 19 & 8.7 & +45.6 & 19 \\
Thermal, GW & 0.06 & +1.99 & 8 & 0.13 & $-$0.52 & 15 \\
Wind, GW & 0.37 & $-$1.69 & 10 & 0.26 & $-$1.67 & 10 \\
Solar, GW & 0.99 & $-$2.26 & 8 & 0.73 & +2.22 & 10 \\
Storage, GW & 0.96 & $-$0.96 & 1 & 1.01 & $-$1.01 & 1 \\
\hline
\multicolumn{7}{|l|}{\textbf{Zonal generation}} \\
\hline
Total, \% & 6.0 & $-$40.9 & 20 & 4.8 & +32.9 & 19 \\
Thermal, TWh & 0.53 & +5.21 & 8 & 0.39 & +2.95 & 8 \\
Wind, TWh & 1.32 & $-$6.27 & 10 & 0.90 & $-$6.19 & 10 \\
Solar, TWh & 1.88 & +4.43 & 10 & 1.60 & +4.51 & 10 \\
Storage, TWh & 0.20 & $-$0.20 & 1 & 0.15 & $-$0.15 & 1 \\
\hline
\multicolumn{7}{|l|}{\textbf{Per line}} \\
\hline
New capacity, GVA & 0.00 & +0.32 & 16–20 & 0.00 & +0.24 & 17–5 \\
Hours $>$80\%, pp & 11.0 & +45.5 & 17–5 & 0.5 & +18.1 & 13–4 \\
\hline
\end{tabular}
\end{minipage}
\end{table}

\textbf{Transport and DC reroute power through the network.}

\begin{figure}[p]
    \centering
    \includegraphics[width=\linewidth]{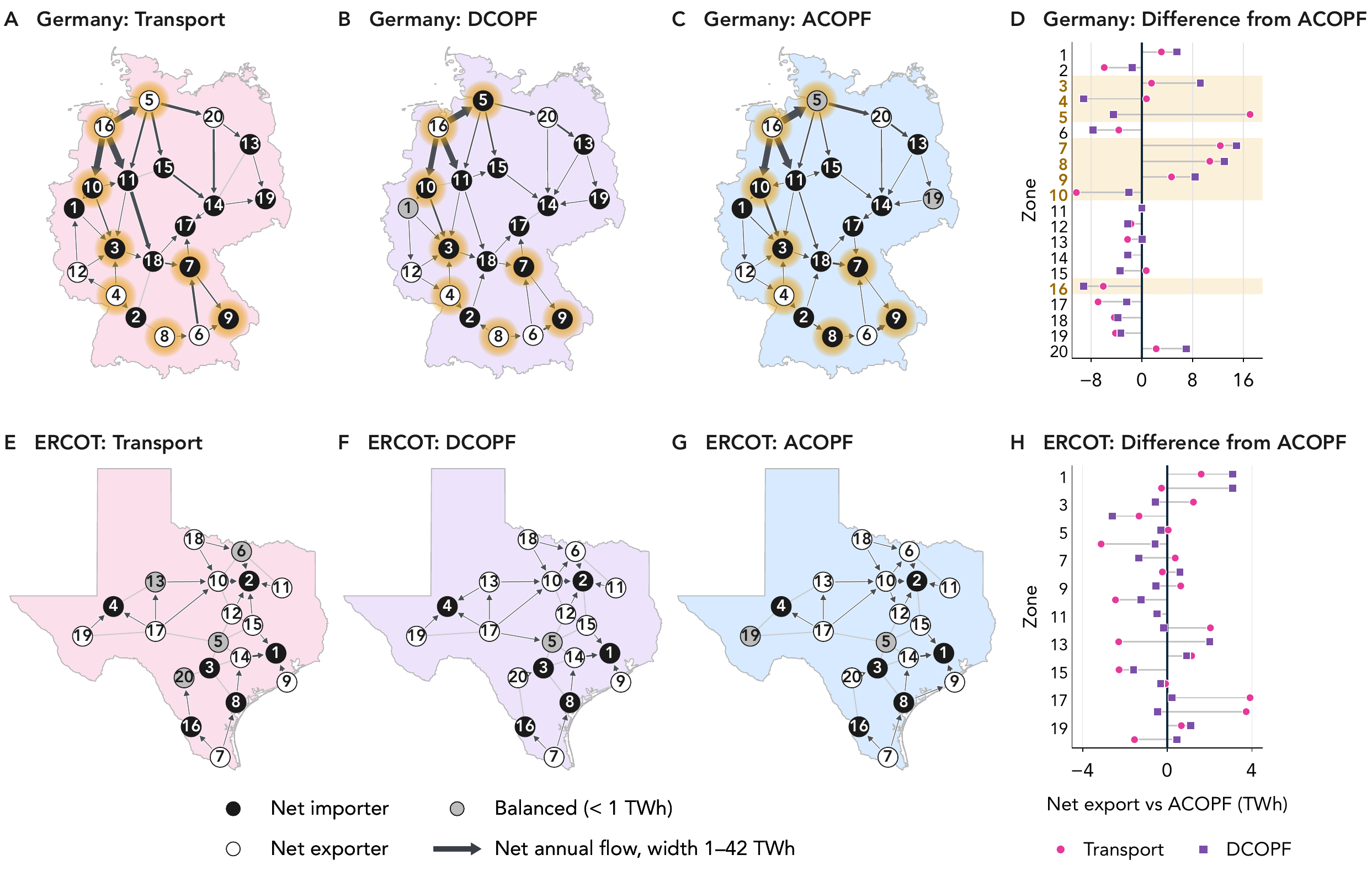}
    \caption{\textbf{Net annual power flows and zonal import/export roles by formulation.} (A--C) Germany and (E--G) ERCOT under transport, DCOPF and ACOPF. Arrows show the net annual flow on each line, with width proportional to energy (1--42\,TWh, one scale for all maps); lines with less than 1\,TWh of net flow carry no arrow. Net annual flow understates the use of lines that carry power in both directions alternately throughout the year. Zone fill indicates whether that zone is: net importer (black), exporter (white), or balanced within 1\,TWh (grey). (D, H) Net annual export of each zone under transport and DCOPF minus that under ACOPF (zero line). Gold zone highlights (A--C) and rows (D) mark zones where either approximation differs from ACOPF by more than 8\,TWh; no ERCOT zone meets this criterion.}
    \label{fig:flows}
\end{figure}

Because demand is identical across transmission formulations, differences in capacity siting change zonal imports and exports, and consequently, power flows through the network. Fig.~\ref{fig:flows} traces the consequence of the siting differences in Fig.~\ref{fig:zone_bars} through the network. Germany is supply-driven: offshore wind in zone 16 exports 103--112\,TWh per year, around a fifth of national demand, toward importing load centers in the west and south. ERCOT is demand-driven: in the modeled year, Dallas (zone 2) and Houston (zone 1) account for 43\% of ERCOT's demand and together import around 60\,TWh, approximately a third of their combined annual demand, from surrounding exporting zones. Comparing formulations in Germany, eight of the 20 zones change their annual net position by more than 8~TWh under transport or DC relative to ACOPF, and two of them (zones 5 and 8) switch between net importer and net exporter. Summed over lines, transport routes 46\% of ACOPF's net annual line flow differently (DCOPF: 29\%). This can be compared with 27\% (DCOPF: 14\%) in ERCOT, where only four lines change by more than 2\,TWh from ACOPF.  

Zone 5 in northern Germany provides an example of how the network representation drives these differences. Under transport, onshore wind in zone 5 and offshore wind in neighboring zone 16 do not compete for transmission capacity, whereas under DCOPF and ACOPF they do. Zone 5 has the highest \textit{onshore} wind capacity factor in the German system, and its wind output is strongly correlated with the \textit{offshore} generation in zone 16 ($r=0.67$). Transport installs 10.7\,GW of onshore wind at zone 5, compared with just 2.0\,GW under DCOPF and 4.2\,GW under ACOPF. Under transport, zone 5 is a 17\,TWh net exporter, with two of the lines carrying its exports at their limits in over a fifth of all hours. Under the DCOPF and ACOPF formulations, the capacity constraints on these lines almost never bind; instead, because power divides across parallel paths according to line impedances, part of zone 5's output flows back through zone 16. In other words, zone 5's generation also uses the lines leaving zone 16, which zone 16's own offshore wind relies on to reach the load centers. These lines already carry the largest flows in the system (Fig.~\ref{fig:flows}A--C) and are above 80\% of their rated capacity in around 40\% of hours under ACOPF. Because the onshore and offshore wind resources are correlated, zone 16's lines are most heavily loaded exactly when zone 5 is producing. Zone 5 therefore competes with the more productive offshore wind in zone 16 for the same line capacity, a competition that the physics-based formulations capture and the transport model cannot see.

\textbf{In Germany, the approximations build less transmission and on different lines.}

\begin{figure}[p]
    \centering
    \includegraphics[width=\linewidth]{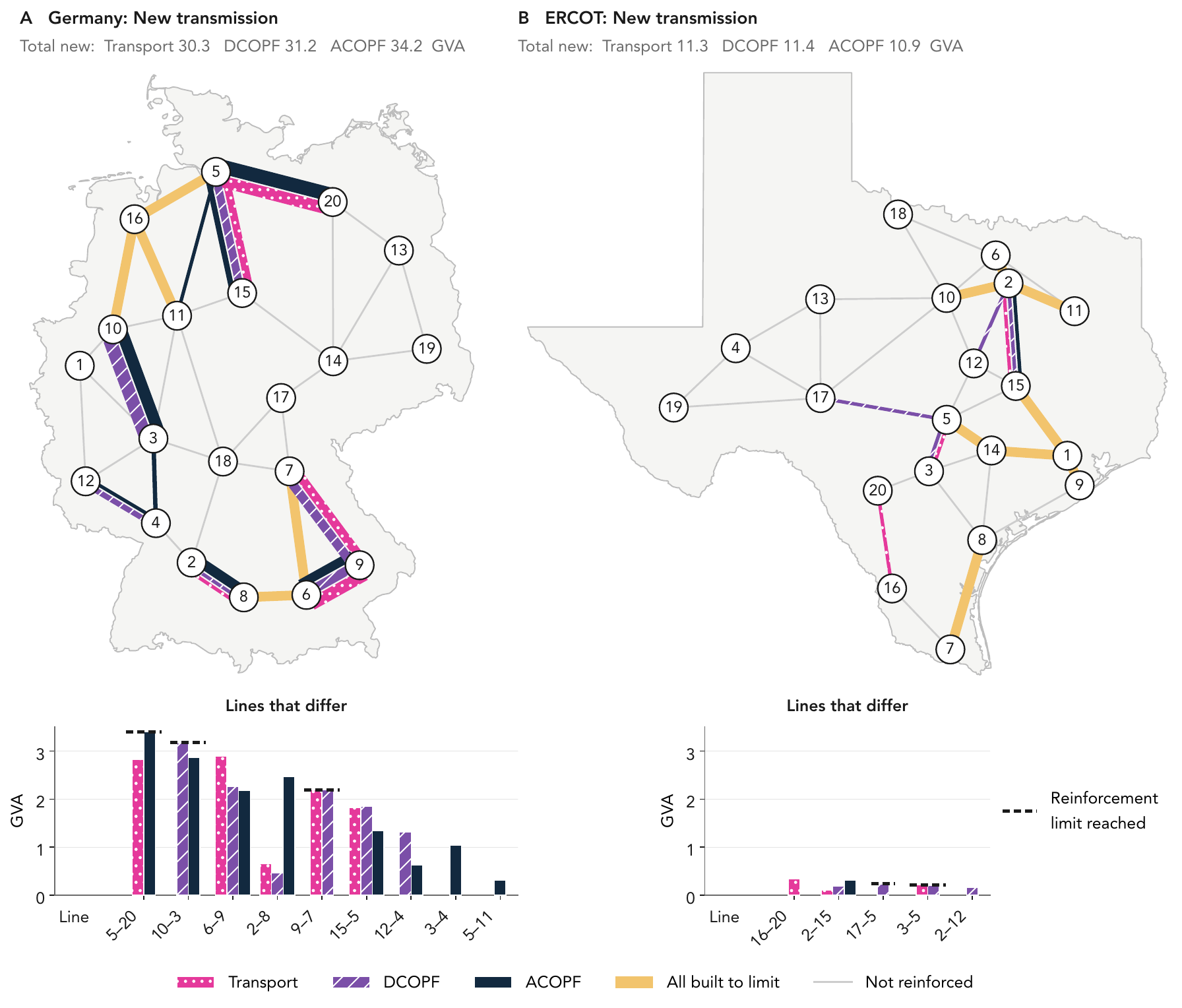}
    \caption{\textbf{New transmission capacity by formulation.} (A) Germany and (B) ERCOT. Gold lines are reinforced to their maximum allowed capacity in all three formulations. Different colored strokes show which lines were reinforced under different formulations; a missing stroke means the line is not reinforced in that formulation. Stroke width is proportional to new capacity, above a minimum width; builds below 0.1\,GVA are not shown. Panels below each map show the amount of new capacity on the lines that differ, with dashes marking a line's reinforcement limit where it is reached. Totals include all lines. All capacities are apparent power 	(GVA; see \hyperref[methods:flow_limits]{Methods, \emph{Flow limits}}).}
    \label{fig:network}
\end{figure}

Since lines are expanded where power flows approach their capacity limits, rerouted flows also change which lines are selected for expansion under each formulation (Fig.~\ref{fig:network}). In Germany, ACOPF installs 34.2\,GVA of new line capacity, transport 30.3\,GVA ($-$11\%), and DCOPF 31.2\,GVA ($-$9\%). The approximations also distribute transmission capacity differently across lines. Of the 14 lines reinforced in any formulation, transport differs from ACOPF on 9 and DCOPF on 8, and 4--5 lines are reinforced in one formulation but not the others. Some of the largest single-line differences occur on lines connecting zones that change in terms of their import/export role (Fig.~\ref{fig:flows}). For instance, ACOPF reinforces the line connecting zone 2 to zone 8 by 2.5\,GVA but transport and DCOPF just 0.7 and 0.5\,GVA, consistent with zone 8 importing under ACOPF but exporting under both approximations. In ERCOT, the total installations are more similar across formulations, and only 3--4 out of 13 reinforced lines differ from ACOPF, none by more than 0.5\,GVA. In ERCOT, many reinforced lines are built to their maximum across all formulations, so the reinforcement limits rather than the power-flow representation largely determine which lines get built. 

While a potential explanation for ACOPF's additional transmission in Germany would be that reactive flows occupy line capacity, this is not the case for these particular model outputs. When an AC line is above 80\% of its rating, its flow-weighted power factor rises to 0.99 in Germany and 0.97 in ERCOT, so congested lines carry almost entirely active power. We also find that German transmission is not sized for peak demand. No AC line exceeds 81\% of its rating on the day of peak demand. Instead, the most heavily used lines are those leaving the offshore wind hub in the north, which are above 80\% of their rating in around 40\% of all hours. Observed transmission differences arise from how active power flows through the network, which shapes where generation is sited (as we saw for zone 5) and, in turn, which lines are reinforced.

\textbf{Storage and OCGT generate reactive power disproportionately.}
\begin{figure}[htbp]
    \centering
    \includegraphics[width=\linewidth]{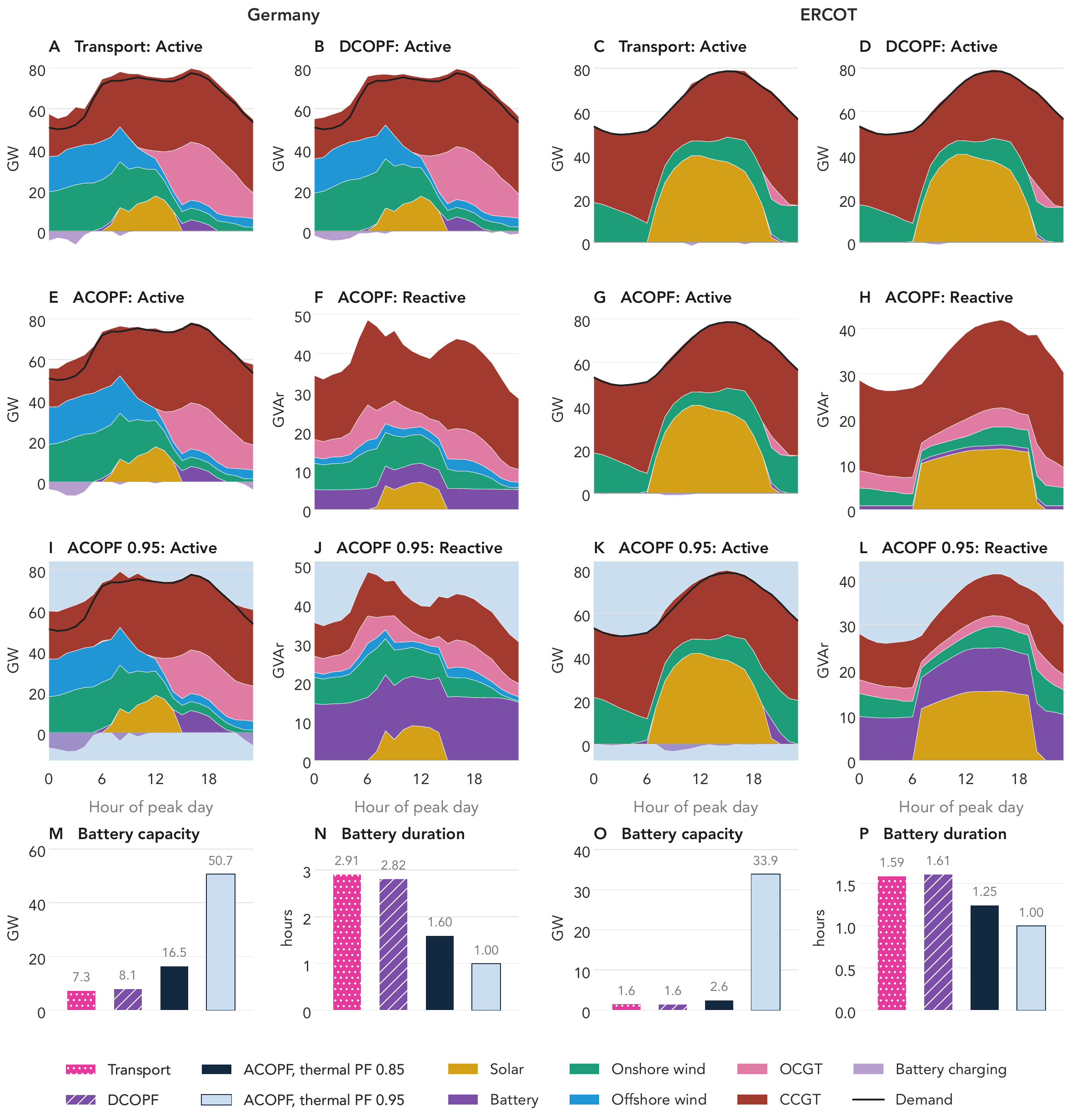}
\caption{\textbf{Peak-demand dispatch and battery investment by formulation.} Germany is represented in the left two columns and ERCOT under the 5\,GW per-zone thermal cap the right. (A--D) Active power dispatch under transport and DCOPF on the day containing the year's peak demand hour. (E--H) Active and reactive power dispatch under ACOPF on the same day. (I--L) The same for an ACOPF case in which the power factor of all thermal generation is tightened from 0.85 to 0.95 to match inverter-based resources, with background color matching the 0.95 AC battery charts for emphasis. Battery charging is drawn below the 0\,GW line in the active power plots. Where generation exceeds demand, the difference is transmission losses and storage charging. (M, O) Installed battery power capacity and (N, P) mean battery duration (energy over power capacity) for the three formulations and ACOPF 0.95. }
    \label{fig:pq_dispatch}
\end{figure}

Although reactive power plays little role in line congestion, the AC formulation must still balance reactive demand in every zone and hour. In our experiments, the resources supplying reactive power deviate from those supplying active power. Total annual power generation only differs by around 1\% across the three formulations in both systems (Table~\ref{tab:headline}). The active-power panels of Fig.~\ref{fig:pq_dispatch} (A--E, G) show this agreement also holds hour-to-hour on the peak-demand day (defined as the day that features the highest hour of active demand across representative weeks), where we see active dispatch is similar under all three formulations. For reactive power dispatch under ACOPF (F, H), however, the narrative is very different. 

The mix of resources that serves reactive demand features strong differences from the active mix. In Germany, batteries supply 13.0\% of reactive output on the peak-demand day while serving only 1.7\% of active output. Over that day, they provide 4.3 units of reactive power for every unit of active power they discharge (the full-year weighted ratio is the same). ERCOT shows the same pattern more strongly, with 5.4 on the peak-demand day and 8.2 across the year. The gap between these ratios is explained by ERCOT's batteries being idle on active power (defined as their aggregate output being below 0.1\% of installed capacity) for 88\% of modeled hours, while continuing to supply reactive power. The peak-demand day is one of the few exceptions when they are active because of the system's high demand, yielding a lower reactive-to-active ratio. Germany's storage undergoes more frequent charge–discharge cycles but exhibits the same behavior as ERCOT, being idle in 59\% of hours. As we discuss further in the next section and depict in Fig.~\ref{fig:pq_dispatch}, models that only track active power are insensitive to the role of storage in reactive support.

OCGTs are also more prominent in reactive than active power support. Across the weighted year, ERCOT's OCGT fleet supplies 10.3\% of the system's reactive output compared to only 0.8\% of its active output. Similarly, Germany's supplies 10.7\% versus 1.8\%. These ratios depend on how generator reactive capability is modeled. Our baseline reactive power constraint allows synchronous plants to provide more reactive power than inverter-based resources (IBRs), such as renewables and battery storage, while limiting reactive power provision when these resources are operating at very low output (see \hyperref[methods:qlimits]{Methods, \emph{Reactive power capacity limits}}). In \ref{si:qlimits}, we test two additional variants on that modeling choice. In the first, IBRs can provide reactive power even when they are not generating active power. In the second, reactive power capability depends on current active generation rather than installed capacity, so idle storage does not contribute reactive power (see Fig.~\ref{fig:q_envelope}).

\textbf{ACOPF builds more battery capacity, at shorter duration, to serve reactive demand.}

In both Germany and ERCOT, battery capacity diverges strongly between transmission formulations. ACOPF builds 1.6--2.3$\times$ more battery storage than either approximation in order to provide the reactive power support discussed previously. We conduct a sensitivity analysis in which the reactive supply limit of thermal generators is tightened to match the limit of IBRs (Fig.~\ref{fig:pq_dispatch}). Removing the thermal fleet's advantage in reactive generation causes the system to substitute battery inverters for synchronous reactive support. Installed battery capacity rises sharply under the tightened limits (Germany: 16.6 to 50.7 GW; ERCOT: 2.6 to 33.9 GW), while duration (defined as installed energy capacity divided by power capacity) falls to the 1-hour minimum in both systems, indicating that batteries are built primarily for power rather than energy capacity. In our reactive supply limits, reactive capability scales with a battery's power capacity and not with its stored energy. The system supplies reactive support by installing more power capacity, buying only the energy capacity needed to meet the minimum-duration constraint. Batteries in the ACOPF case are providing reactive power support alongside energy arbitrage, with reactive support becoming even more dominant under the tighter reactive generation limit for thermal generators. Transport and DCOPF cannot represent this service at all, as reactive power is not represented in either model. 

\textbf{DCOPF and ACOPF take far longer than transport to solve.}

All formulations solve the same 20-zone system with transmission expansion, using 20 representative periods for Germany and 32 for ERCOT. A linear optimization problem, transport solves in a few minutes with Gurobi in both systems. ACOPF is a nonlinear and nonconvex optimization problem. DCOPF is linear, but in the expansion setting, the parallel-circuit transmission expansion modeling (see \hyperref[methods:compute]{Methods, \emph{Computational setup}}) adds a nonlinear network-expansion term to both the DCOPF and ACOPF models. Because of this nonlinearity, we use Ipopt to solve both DCOPF and ACOPF. DCOPF solve times are roughly two orders of magnitude greater than transport, and solving ACOPF adds approximately another factor of 2 to the runtime (Table~\ref{tab:solve_times}). On a smaller system, \citet{weng2026modeling} found that ACOPF runtimes grow steeply with the number of representative weeks while transport runtimes barely change, and in ERCOT, increasing from 20 to 32 periods raised the ACOPF solve time from 6.0 to 10.1 hours (compare Tables~\ref{tab:solve_times}, \ref{tabS:nonx}).

\begin{table}[htbp]
\centering
\caption{\textbf{Solve times by power-flow formulation and solver.} Wall-clock times on 32 cores of an AMD EPYC 9474F node (MIT ORCD Engaging cluster); see \hyperref[methods:compute]{Methods, \emph{Computational setup}}.}
\label{tab:solve_times}

\begin{minipage}[t]{0.48\linewidth}
\scriptsize
\setlength{\tabcolsep}{3pt}
\centering
\textbf{Germany}

\medskip
\begin{tabular}{|l | l | r | r|}
\hline
\textbf{Formulation} & \textbf{Solver} & \textbf{Problem type} & \textbf{Solve time} \\ [1ex]
\hline
Transport & Gurobi & LP            & 126 s ($\approx$2 min) \\ [1ex]
\hline
DCOPF & Ipopt  & NLP           & 10{,}145 s ($\approx$2.8 h) \\ [1ex]
\hline
ACOPF & Ipopt  & NLP           & 18{,}040 s ($\approx$5.0 h) \\ [1ex]
\hline
\end{tabular}
\end{minipage}
\hfill
\begin{minipage}[t]{0.48\linewidth}
\scriptsize
\setlength{\tabcolsep}{3pt}
\centering
\textbf{Texas (ERCOT), 5\,GW per-zone thermal cap}

\medskip
\begin{tabular}{|l | l | r | r|}
\hline
\textbf{Formulation} & \textbf{Solver} & \textbf{Problem type} & \textbf{Solve time} \\ [1ex]
\hline
Transport & Gurobi & LP            & 200 s ($\approx$3 min) \\ [1ex]
\hline
DCOPF & Ipopt  & NLP           & 17{,}814 s ($\approx$4.9 h) \\ [1ex]
\hline
ACOPF & Ipopt  & NLP           & 36{,}414 s ($\approx$10.1 h) \\ [1ex]
\hline
\end{tabular}
\end{minipage}
\end{table}

\section*{DISCUSSION}

Given the urgency of decarbonizing the power sector and rapid developments in both supply and demand, reliable and transparent power system modeling is essential. CEM results are known to be sensitive both to uncertain inputs, such as technology costs and weather years,\cite{shirizadeh2022sensitive, pecora2025quantifying} and to how the system is represented, such as its temporal and spatial resolution.\cite{schyska2021sensitivity,  frysztacki2021strong, serpe2025importance, qiu2024decarbonized} How the physics of the transmission network is modeled is a representation choice that has been studied far less.\cite{neumann2022assessments, khorramfar2026optimization} Here, we incorporate full ACOPF into a CEM that co-optimizes generation, storage and transmission. Using 20-zone models of Germany and ERCOT, we assess where and when a full AC representation changes planning decisions, and whether the mechanisms behind those changes could be identified or captured more tractably.

\textbf{Implications for planning.} Across both systems, similar aggregate costs, emissions, and generation capacities conceal substantial differences in where capacity is built and how the network is used (Tables~\ref{tab:headline} and \ref{tab:zonal_stats}). Broad investment patterns follow resource availability, but zonal amounts differ, sometimes by enough to change whether a region imports or exports electricity (Fig.~\ref{fig:flows}, Table~\ref{tabS:zonal_balance}). These zonal differences are not incidental, tracing back to identifiable mechanisms of network congestion and reactive power support. Agreement on system-wide totals is therefore not, on its own, sufficient evidence that a simplified representation of transmission produces reliable zonal investment decisions. Nor are small aggregate differences always negligible once they are valued. The approximations emit only a few percent more than ACOPF, but because the models do not price carbon, this difference does not appear in total system cost. Applying a social cost of carbon of about \$185 per metric ton,\cite{rennert2022comprehensive} these additional emissions are comparable to or larger than the differences in total system cost, and under DCOPF they exceed the entire network expansion investment.

A reassuring tautology emerges from the pattern of spatial differences we observe: transmission representation matters where transmission matters. This observation shows up both at the local and broader system levels. Locally, zone 5 in Germany illustrates the mechanism. Abundant onshore wind encourages investment under transport, but its output is correlated with offshore wind in neighboring zone 16, and physics-based routing makes the two compete for shared transmission capacity (Fig.~\ref{fig:flows}). More broadly, differences in generation siting and transmission reinforcement are larger in Germany than in ERCOT, which points to the spatial separation of supply and demand driving differences between transmission representations. Germany's wind resources are concentrated far from its largest demand centers, whereas thermal generation in ERCOT can be located near them. We conduct a sensitivity analysis of zonal thermal capacity caps in ERCOT that supports this interpretation (see \ref{si:caps}). Limiting thermal capacity per zone pushes supply away from demand centers and roughly doubles the deviation between CEM outputs under the different transmission formulations (Table~\ref{tabS:caps_vre}). Because the caps also change the generation mix, renewable share, and reactive capability, this comparison does not isolate the effect of geographic separation. Nevertheless, the results suggest that a higher-fidelity network representation becomes more important as supply and demand become more spatially separated. This is particularly relevant in systems with high shares of renewable generation, where resource availability and land constraints can concentrate generation far from load centers. 

Transmission expansion reflects these interactions between generation siting and network use. In Germany, heavily loaded lines carry predominantly active power, and the largest line reinforcements serve bulk wind exports rather than peak demand. In our results, active-power routing drives the reinforcement differences between representations, yet this does not preclude the existence of systems where reactive power flows have a large effect on transmission investment. In ERCOT, most new transmission is built on corridors that every representation reinforces to its limit, largely around Houston and Dallas, so expansion differs little across representations. In a recent comparison of AC and DC transmission expansion, \citet{khanpour2026empirical} add discrete candidate lines to 5- and 9-bus test systems and find that the two formulations yield similar expansion plans and nearly identical costs under standard parameters, diverging mainly when line parameters are varied to exaggerate losses or reactive effects. In our 20-zone systems, we instead reinforce existing corridors continuously and co-optimize renewables and storage with transmission investment. As in their study, we find that total system costs agree across formulations; unlike their study, we find that siting and reinforcement differ even under standard assumptions where the network constrains the system. This difference reflects the interactions between generation and transmission investment that their setting does not capture. 

Reactive power requirements also shape investment decisions. The resources providing reactive power can differ substantially from those serving active power demand. Battery storage and OCGT generators that appear underutilized from an active power perspective provide substantial reactive support (Fig.~\ref{fig:pq_dispatch}), and ACOPF builds more battery power capacity with shorter durations than the simplified representations. The AC formulation values a service the simplified models cannot see. Analyses based on those models may therefore understate its system value and provide an incomplete basis for planning and technology decisions. 

The differences we observe in generation and transmission siting, and in whether zones import or export, can have distributional implications. Recent work links such differences to distributional outcomes through both regional electricity prices and the location of new infrastructure.\cite{de2027group} In our analysis, hourly zonal prices, taken as shadow prices of the power-balance constraint, differ little across formulations in either system (Fig.~\ref{figS:price_duration}). The differences that do occur are concentrated in low-price hours and, in Germany, between wind-rich regions and other zones. Actual electricity market prices also depend on market design, congestion within zones, and other conditions that our model does not represent. In ERCOT, for example, wind growth has been shown to shift congestion costs toward distant load centers, and growing large loads in South and West Texas to alter regional price differentials,\cite{madadkhani2026mapping} so the siting differences we observe may have a greater effect on market outcomes than our modeled shadow prices suggest. Prior studies find that the local benefits and burdens of energy infrastructure development are often distributed unevenly across communities.\cite{cranmer2023energy, mueller2020burdened, liljenfeldt2017distributional} Planning models inform expectations of which regions, and potentially which populations, will host new infrastructure. Our results show that this siting can depend on the choice of network representation, and distributional assessments based on planning model outputs inherit that dependence.

\anchor{dis:limitations}\textbf{Limitations of the study.} Representing AC physics in a CEM requires a series of modeling choices, including synthetic reactive demand, reactive capability limits for each resource, shunt compensation for line charging effects, and spatial and temporal aggregation. These choices can affect our results. For example, the additional storage built under ACOPF largely disappears when inverter-based resources may supply reactive power at any output level (Fig.~\ref{fig:q_envelope}). Reactive capability requirements are set by grid codes and differ between systems, so the appropriate assumption depends on the system being modeled. We do not argue that our implementation is the best or only one, and our sensitivity analyses test only some of our choices. 

For reactive power modeling, we did not test the following two modeling extensions. First, simplified models could track reactive power in the same way as active power. Each zone would balance its reactive supply and demand, with limits on how much each resource can supply and each line can carry. Such a linear representation might recover some of the storage investment signal observed under ACOPF. Because it ignores voltage, however, it would remain an approximation, and its accuracy against ACOPF is untested. Second, our resource set allows reactive support only from generators and battery inverters, which is narrower than in reality. STATCOMs, synchronous condensers and switched capacitor banks supply reactive power without the cost of energy storage,\cite{ercot2010crez, chaneliere2022reactive, birchfield2018power} and including them would likely reduce reliance on battery inverters.

We note three further limitations. First, we do not establish at what spatial resolution AC becomes important for planning decisions. Prior work finds that voltage and reactive requirements depend largely on intrazonal conditions, and spatial resolution affects outcomes even under the transport model.\cite{frysztacki2021strong, serpe2025importance} Second, each portfolio is evaluated only within the formulation that produced it, and we did not run AC power flow on the outputs of the simplified models. Given that DCOPF solutions are known to never be AC-feasible,\cite{baker2021dcopf} it is unclear whether the resulting investment portfolios can be operated feasibly when full AC network constraints are considered. Third, capacity expansion problems often have a flat optimum, in which quite different portfolios have similar costs.\cite{neumann2021near} Together with the nonconvexity of ACOPF, this means that some siting differences may reflect near-optimal alternatives or local solutions rather than network physics alone. With that said, a preliminary test shows that the differences we observe between formulations' results cannot be attributed merely to flat optima. When ACOPF in Germany without network expansion is required to build at least the transport capacity portfolio, total cost rises by only 0.8\% relative to unconstrained ACOPF, but battery capacity roughly doubles, with shorter durations consistent with reactive support. The transport portfolio is therefore not sufficient under AC power flow, and ACOPF must add substantial short-duration storage to operate it. Although the two AC solutions cost almost the same, they are not interchangeable, because the transport portfolio lacks the reactive support that AC power flow requires.

\textbf{Outlook.} Our findings point to practical next steps. Transport solves about two orders of magnitude faster than the physics-based expansion formulations (Table~\ref{tab:solve_times}). Rather than embedding full AC physics everywhere, our AC results can guide the design of (near-)linear, AC-inspired models that capture the effects that change planning outcomes. DCOPF shows that using partial network physics already narrows the siting and reinforcement differences, but with network expansion, the optimization problem is nonlinear and only about twice as fast as ACOPF, while omitting reactive support entirely. Targeted linear constraints may offer a better trade-off. Limits on shared export bottlenecks could capture the competition between correlated resources, and the reactive-power representation suggested in \hyperref[dis:limitations]{Limitations} could recover part of the investment signal. A natural next step is testing such constraints against AC solutions across different systems and network resolutions. Simple models serve aggregate planning questions well, while siting, line reinforcement, and reactive-support device decisions in network-constrained regions merit closer attention to the underlying network physics.

\section*{METHODS}

\subsection*{AC Optimal Power Flow}

We integrate AC optimal power flow constraints into GenX, adapting the single time-step formulation from the Power Grid Library\cite{pglib} to operate across the full planning horizon. For each time period ($t \in \mathcal{T}$), zone ($z \in \mathcal{Z}$), generator ($g \in \mathcal{G}$), and transmission line ($l \in \mathcal{L}$), there are constraints which govern complex voltages ($V_{z,t} \in \mathbb{C}$), complex generator dispatch ($S_{g,t} = P_{g,t} + jQ_{g,t}$), and complex line flows ($S_{l,t}, S^{\mathrm{rev}}_{l,t} \in \mathbb{C}$) in both forward and reverse directions.

A reference bus angle is fixed at ($\angle V_{\mathrm{ref},t} = 0$), voltage magnitudes are bounded by upper and lower limits ($\underline{v}_z \leq |V_{z,t}| \leq \overline{v}_z$), and angle differences across each line $l =(i,j)$ are constrained by ($\underline{\Delta\theta}_l \leq \angle V_{i,t} - \angle V_{j,t} \leq \overline{\Delta\theta}_l$). Generator apparent power is bounded by ($\underline{S}_g \leq S_{g,t} \leq \overline{S}_g$), where the active and reactive components are constrained separately.

Forward and reverse complex power flows on each line are given by

\begin{equation}
S_{l,t} = \left(Y_l - \mathbf{j}\tfrac{B^c_l}{2}\right)\frac{|V_{i,t}|^2}{|T_l|^2} - Y_l \frac{V_{i,t} V^*_{j,t}}{T_l},
\label{eq:flow_fwd}
\end{equation}

\begin{equation}
S^\mathrm{rev}_{l,t} = \left(Y_l - \mathbf{j}\tfrac{B^c_l}{2}\right)|V_{j,t}|^2 - Y_l \frac{V^*_{i,t} V_{j,t}}{T^*_l},
\label{eq:flow_rev}
\end{equation}

\noindent where ($Y_l \in \mathbb{C}$) is the series admittance, ($B^c_l \in \mathbb{R}$) the line charging susceptance, ($T_l \in \mathbb{C}$) the complex tap ratio (equal to 1 for non-transformer lines), and $\mathbf{j}$ represents the imaginary unit. Note that, while we have included the possibility of modeling transformers within our GenX implementation, in all of our analyses in this paper, we have set the tap ratios to 1 and phase shift angles to 0, as we lacked transformer data in the systems we analyzed. Apparent power thermal limits are enforced in both directions: ($|S_{l,t}| \leq \overline{S}_l$) and ($|S^\mathrm{rev}_{l,t}| \leq \overline{S}_l$). Complex power balance at each zone is given by

\begin{equation}
\sum_{g \in \mathcal{G}_z} S_{g,t} = \sum_{l \in \mathcal{L}_z} \left(S_{l,t} + S^\mathrm{rev}_{l,t}\right) + D_{z,t} + Y^{\mathrm{sh}}_z |V_{z,t}|^2,
\label{eq:balance}
\end{equation}

\noindent where ($D_{z,t} \in \mathbb{C}$) is complex net demand and ($Y^{\mathrm{sh}}_z \in \mathbb{C}$) is the zonal shunt admittance.

These constraints are coupled to the existing GenX formulation by linking the AC active power generation ($P_{g,t}$) to GenX's generator dispatch variable, the forward active line flow ($\mathrm{Re}(S_{l,t})$) to GenX's net line flow variable, and the sum of forward and reverse active flows to GenX's transmission loss variable, with transport-model loss percentages disabled. Apparent power thermal limits use GenX's available transmission capacity, which evolves with investment decisions when network expansion is enabled. The objective function retains GenX's original linear operational cost structure; no quadratic or piecewise linear dispatch cost is introduced.

\subsection*{DC Optimal Power Flow}

For DCOPF, we use GenX's existing formulation.\cite{genx} It applies the standard linear approximation of AC power flow, in which voltage magnitudes are fixed at 1 per unit, angle differences across lines are assumed small, line resistance is neglected relative to reactance, and reactive power is ignored. Each zone is assigned a voltage angle $\angle V_{z,t}$, and the active power flow on each line $l=(i,j)$ is a linear function of the angle difference across it,

\begin{equation}
P_{l,t} = b_l \left(\angle V_{i,t} - \angle V_{j,t}\right),
\label{eq:dc_flow}
\end{equation}

\noindent where $b_l$ is the line susceptance (which, in settings with no resistance, is the inverse of reactance). As in the AC case, the reference bus angle is fixed at zero and angle differences are bounded. Flows are subject to GenX's existing active power line limits and zonal power balance, so DCOPF differs from the transport model only through Eq.~\ref{eq:dc_flow}, which determines how flows distribute across the network. Extensions to this formulation for network expansion and line losses are described under Modeling Details.

\subsection*{Modeling Details}

\anchor{methods:qlimits}\textbf{Reactive power capacity limits.} Each generator's reactive power output is bounded by a symmetric box constraint sized from a minimum rated power factor, $Q^{\max}_y = \tan(\arccos(\mathrm{PF}_y)) \cdot P^{\mathrm{cap}}_y, Q^{\min}_y = -Q^{\max}_y$, in place of a full capability curve. In our baseline, synchronous generators use $\mathrm{PF}=0.85$; inverter-based resources (wind, solar, storage) use the tighter $\mathrm{PF}=0.95$, consistent with typical inverter nameplate ratings and grid-code requirements\cite{britto2026building, vde4120, sarkar2018reactive}. In \ref{si:pfsweep} and Fig.~\ref{fig:pq_dispatch}, we perform a sensitivity analysis where we vary the synchronous generator limit to also have $\mathrm{PF}=0.95$, matching the ratings of inverter-based devices. For VRE specifically, we apply a dead-zone following German grid codes:\cite{vde4120} no reactive power capability below 5\% active power availability, ramping linearly to the full $\mathrm{PF}=0.95$ capability level at 20\% availability.

\anchor{methods:nx}\textbf{Network expansion.} Transmission network expansion under the transport model is straightforwardly linear: line capacity is a continuous investment variable that does not affect network parameters, which are ignored by the transport model. However, transmission expansion is more complex for both AC and DC power flow-based transmission. Electricity network parameters (susceptance for the DC case, full admittance for AC) scale with the new installed line capacity and must be linked to the investment decisions. We follow \citet{brown2017pypsa} and parametrize expansion through the number of parallel circuits $n_l$. The effective admittance is then $Y_l^\mathrm{eff} = n_l Y_l^0$, where $Y_l^0$ is the per-circuit admittance. We relax the integrality of $n_l$ and treat it as a continuous variable in both the AC and DC cases, and the thermal line rating scales proportionally. In the DC case, the effective susceptance multiplies the voltage angle difference, introducing a bilinear term that renders the DC network expansion problem nonlinear. The AC case introduces additional nonlinearity: both the series admittance and the line charging are functions of $n_l$, and these multiply nonlinear voltage terms in the flow calculations.

While we adopt the continuous parallel-circuit scaling approach, it is not the only viable formulation. Alternatives include expanding thermal ratings while holding admittances fixed,\cite{khanpour2026empirical} using integer circuit counts for physical realism at the cost of a mixed-integer nonlinear program, or allowing investment in qualitatively different line types (e.g., higher-voltage AC) with distinct line parameters. 

\textbf{Reference bus and voltages.} AC power flow requires a slack bus at which the voltage angle is fixed to zero; all other angles are measured relative to this reference. PyPSA includes a slack zone in its data that we set to be our reference. For voltage magnitude initialization, we set $|V_{z,0}| = V_z^\mathrm{nom}$ using nominal voltage levels from the network data, which provides Ipopt with a physically meaningful starting point. We set all zonal voltage magnitude limits to $\pm$5\% of the (uniform) line voltages across the system according to ERCOT's Nodal Operating Guide.\cite{ercot2026nog} Voltage angle differences between adjacent zones are limited to 30$^\circ$.

\anchor{methods:flow_limits}\textbf{Flow limits.} Line ratings in our source data are apparent-power (MVA) capacities. The AC formulation can use these numbers directly. Transport and DC do not model reactive power, so the same figure would get applied as a pure active-power limit, overstating what the line could actually deliver. To make the three formulations comparable, we scale both the thermal rating and the network reinforcement costs in the transport and DC cases by a power factor of 0.9, the mean demand power factor across the network. This is an approximation: a single system-wide power factor cannot fully capture the spatiotemporal variance in reactive power loading. Line ratings are multiplied by this factor, so that a given 100 MVA line offers 90 MW active-power transfer capacity. Line reinforcement costs are divided by the same power factor, so that reinforcing a line costs approximately 11\% per MW of real power more than per MVA of apparent power. These adjustments are applied to all transport and DC runs, with AC unscaled. To compare installed transmission on a common basis, we report capacity for all formulations in apparent power (GVA), converting transport and DC capacities back using this same factor. 

\textbf{Line losses.} The transport model allows for linear line losses as a percentage of flow through a line. DCOPF is typically lossless, but for closer consistency with AC, our baseline is a \textit{lossy DC} model where we apply line losses in the same manner as for transport. For the line loss percentage, we use a figure of 3.5\% across all lines for Germany, based on the flow-weighted mean line losses in the top 10\% most-utilized lines observed in our AC runs. For ERCOT, congestion is spread more evenly, resulting in a lower uniform loss of 2.5\% calculated in the same manner. 

\textbf{Shunt reactors.} In the AC case, we add shunt reactors to account for the capacitive effects of line charging on reactive power balance. Based on reactive power studies of Germany and ERCOT,\cite{brueckl2021blindleistung, ercot2010crez} we add a shunt reactor to each zone, with magnitude equal to 100\% of the line charging susceptance $B^c_l/2$ at the ends of line $l$ to eliminate the effects of line charging.

\anchor{methods:compute}\textbf{Computational setup.} 
All models are solved on the MIT Office of Research Computing and Data (ORCD) Engaging cluster, on dual-socket nodes with two 48-core AMD EPYC 9474F processors (3.6\,GHz base clock) and 384\,GB of RAM. Solve times are reported for jobs allocated with 32 CPUs and 48\,GB of memory. The single exception was ERCOT's 32-period ACOPF case, where we doubled the allocation to 96\,GB due to an out-of-memory error. Both systems use the same solver settings. ACOPF and DCOPF are solved with Ipopt 3.14.19 (via Ipopt.jl 1.13.0) at a convergence tolerance of $10^{-6}$, an adaptive barrier strategy, and a 5{,}000-iteration limit, using the HSL MA86 sparse symmetric-indefinite linear solver with MC64 scaling. Transport is solved with Gurobi 11.0.3 (via Gurobi.jl 1.7.6) using the barrier method, barrier convergence tolerance $10^{-8}$, feasibility and optimality tolerances of $10^{-5}$, and crossover enabled. Gurobi's thread pool is pinned to 32 to match the CPU allocation for each job. All models are built with JuMP 1.29.3 in Julia 1.11.7.

\subsection*{Data collection and pre-processing}\anchor{methods:data} 

\anchor{methods:de}\textbf{Germany.} 
Our model builds on the German electricity network developed by Xiong et al.\cite{xiong2025modelling} We use PyPSA-Eur\cite{PyPSAEur, brown2017pypsa} to extract data and technical specifications for power plants, transmission lines, and electricity demand, including the corresponding time series, and cluster the network to 20 zones with PyPSA-Eur's default load-weighted k-means clustering of bus locations. We include generation from on- and off-shore wind; solar PV, including with horizontal single-axis solar tracker; OCGT; and CCGT. We \emph{exclude} run-of-river, geothermal, and oil due to their small contribution to the German generation mix;\cite{energycharts2024} nuclear, following its completed phase-out in 2023; and coal and lignite, given their planned decommissioning under German energy policy.\cite{kvbg2020} Renewable generation and demand time series are taken from calendar year 2023, the latest year available in PyPSA-Eur at the time of data collection.

Since the input format required by GenX differs from the output format of PyPSA-Eur, we convert the data as necessary (see \hyperref[sec:availability]{Data and code availability}). Investment costs and fixed and variable operation and maintenance (O\&M) costs are taken from the Energy System Technology Data available through PyPSA, using projected costs for 2030. All monetary values are converted to 2023 euros using HICP annual data from Eurostat,\cite{eurostat_hicp_aind} and a discount rate of 7\% is applied following Hofmann et al.\cite{hofmann2025h2} Battery costs are calculated based on Danish Energy Agency's \textit{Technology Data for Energy Storage} for utility-scale lithium-ion batteries for year 2030\cite{dea_storage_technology_data}, which we convert to 2023 euros using the relevant HICP index.\cite{eurostat_hicp_aind} Natural gas and biomass prices are assumed fixed throughout the planning period at 40.88 EUR/MWh and 26 EUR/MWh, respectively. The natural gas price corresponds to the 2023 annual average of the European Gas Spot Index for THE.\cite{bnetza_monitoring2024} For biomass, market price data are limited, so we use the reference assumption from the industry report by \citet{hogan2010biomass}.

\anchor{methods:ercot}\textbf{ERCOT.}
For ERCOT, we build on PyPSA-USA,\cite{Tehranchi_PyPSA-USA_An_Open-Source} replacing some of its default inputs to better suit a 20-zone AC-enabled system. Renewable generation and demand time series are for calendar year 2022, the latest year available in PyPSA-USA and the closest to Germany's 2023. We include generation from onshore wind; solar PV with horizontal single-axis tracking, which made up 94\% of U.S. utility-scale PV additions in 2022;\cite{bolinger2023utility} OCGT; CCGT; and coal. We \emph{exclude} hydropower, which supplied under 0.4\% of ERCOT's generation in 2022,\cite{ercot_fuelmix} and, unlike Germany, do not include offshore wind or biomass, neither of which is a significant source of generation in ERCOT. We also exclude nuclear, because new conventional nuclear capacity is unrealistic in a single-year greenfield build given its long construction lead times.\cite{portugal2018better} Recent advanced nuclear technologies, such as small modular reactors, aim to shorten construction timelines,\cite{doe2024liftoffnuclear} and we leave their inclusion to future work.

The default zonal option from the ReEDS model of the National Lab of the Rockies (NLR),\cite{REEDS, brown2023general} which only defines ERCOT's transfer limits between a few zones, is too coarse to parametrize our 20 zones. The nodal option, the Breakthrough Energy synthetic network of the U.S.,\cite{xu2020us, birchfield2017grid} represents ERCOT with a 2{,}000-bus synthetic case whose voltage levels do not match the real system's, and whose line capacities were raised to eliminate load shedding and then recalibrated to the ReEDS limits. Instead we use Texas A\&M University's \textit{Texas7k} synthetic network of 6{,}717 buses,\cite{tamu7k, texas7k_data} whose voltage levels match ERCOT's, so its lines can be checked against in-service transmission data from HIFLD.\cite{hifld_transmission} We cluster its buses into 20 zones using PyPSA's hierarchical agglomerative clustering approach with Ward linkage,\cite{ward1963hierarchical, brown2017pypsa} grouping buses by their hourly 2022 wind and solar capacity factors from ERA5\cite{hersbach2023era5, hofmann2021atlite} and by whether their lines geographically match real HIFLD lines, so that zones are not joined across ties that exist only in the synthetic case. 

We derive line parameters from Texas7k's own branch ratings and impedances. We reduce the network with a border-zone-preserving Kron reduction to get the impedances.\cite{ploussard2018efficient} For the line transfer capacities, we implement the line-limit-preserving equivalents method from \citet{jang2013line}. The resulting 34 lines have transfer capacities of 221--3{,}608\,MW (median 1{,}129\,MW), comparable to published ERCOT inter-zonal capacities.\cite{kassel2025method} The reduced 20-zone, 34-line network is spatially consistent with published reduced networks validated on real ERCOT data.\cite{martin2017reduced, johnson2021economic}

In addition to the transmission network, we modify the active electricity demand from the default PyPSA-USA output. To allocate electric load to each node, PyPSA-USA distributes the aggregate ERCOT-wide demand time series from EIA according to the normalized active power demand ($P_d$) of the underlying base network, which is spatially distributed based on regional population and assumed per-capita electricity consumption.\cite{texas7k_data} All nodes inherit the same temporal demand profile, differing only in magnitude, and sub-regional variation in load shapes arising from differences in weather exposure or economic activity (e.g., industrial oil and gas activity in West Texas versus residential demand in urban areas) is not captured. To account for this spatial variability and its consequences for transmission and generation needs, we replace the default PyPSA-USA demand with ERCOT's metered hourly demand for its eight weather zones,\cite{ercot_load_hist} which sets both the magnitude and the shape of demand in each weather zone. We then allocate each weather zone's demand across our 20 model zones using county-level shares from NLR's hourly county demand dataset:\cite{obika2026estimated, ercot_county} each county is assigned to the weather zone covering most of its area, weighted by its share of that weather zone's annual energy, and split across model zones by the $P_d$ of the Texas7k buses it contains. Although the same limitation in spatial load allocation remains in the German dataset, we leave the PyPSA-Eur demand representation unchanged for the German test case due to data availability.

Technology costs are from NLR's 2022 Annual Technology Baseline (ATB).\cite{nrel_atb2022} Fuel prices are from EIA data for Texas;\cite{eia_texas_fuel} natural gas uses 2022's monthly price shape, rescaled to the 2022--2025 mean of 3.45\,\$/MMBtu because 2022 prices were an outlier, and coal is held at a constant 1.82\,\$/MMBtu. Battery costs are for utility-scale lithium-ion battery storage. We fit a linear relationship across the ATB's 2--10-hour systems to separate power-related (\$/kW) and energy-related (\$/kWh) costs, and annualize both over a 20-year cost recovery period. Power and energy capacity are sized independently, with duration constrained to 1--8\,h; the upper bound is from PyPSA-USA's longest default battery option.

\textbf{Reactive power demand.}
Reactive power demand is typically not reported for electricity systems. As such, for both Germany and ERCOT, we generate a synthetic reactive demand time series by drawing power factors uniformly within $\pm0.02$ of a mean of 0.9, which we apply to our active demand time series (with the same fixed seed used in each solve). The mean power factor of 0.9 corresponds to the widest reactive power range permitted for transmission-connected demand under EU regulation,\cite{eu2016dcc} and the resulting hourly range of 0.88--0.92 includes the 0.92 power factor assumed in recent open-data U.S. AC power flow models.\cite{britto2026building} In \ref{si:pfsweep}, we also study the effects of choosing a lower mean power factor to represent industrial demand, creating more reactive demand on the system (Fig.~\ref{fig:q_sweep}).

\newpage

\section*{RESOURCE AVAILABILITY}

\subsection*{Lead contact}

Requests for further information and resources should be directed to and will be fulfilled by the lead contact, Lara Booth (lcbooth@mit.edu).

\subsection*{Data and code availability}\anchor{sec:availability}

\begin{itemize}
    \item This paper analyzes existing, publicly available data. Because several sources require users to register or accept license terms before download, we do not redistribute them. The code repository lists each data source with download instructions.
    \item All original code is publicly available at \href{https://github.com/lcbooth/GenX-ACOPF.jl}{github.com/lcbooth/GenX-ACOPF.jl} as of the date of publication. This includes our extension of GenX\cite{genx} with AC optimal power flow, distributed under GenX's GNU General Public License v2.0, and the scripts that build both systems' input data.
    \item For Germany, the network is derived from the prebuilt PyPSA-Eur network\cite{xiong_2024, xiong2025modelling} and contains information from OpenStreetMap ({\copyright}~OpenStreetMap contributors), made available under the Open Database License (ODbL) v1.0. Power plant, demand, and renewable time series are extracted with PyPSA-Eur,\cite{PyPSAEur} using wind data from ERA5\cite{hersbach2023era5} and solar data from SARAH-3 from EUMETSAT's Satellite Application Facility on Climate Monitoring (CM SAF), {\copyright}~EUMETSAT. Technology costs are from PyPSA-Eur\cite{PyPSAEur} with battery costs from the Danish Energy Agency,\cite{dea_storage_technology_data} all converted to 2023 euros using Eurostat HICP data.\cite{eurostat_hicp_aind} Natural gas prices are from the Bundesnetzagentur,\cite{bnetza_monitoring2024} and biomass prices from \citet{hogan2010biomass}.
    \item For ERCOT, the network is based on the Texas A\&M University Texas7k synthetic grid,\cite{tamu7k, texas7k_data} with line routes checked against HIFLD transmission data.\cite{hifld_transmission} Demand is ERCOT's metered hourly load\cite{ercot_load_hist} allocated using NLR's county demand profiles.\cite{obika2026estimated} Hourly wind and solar capacity factors are computed from ERA5\cite{hersbach2023era5} using atlite.\cite{hofmann2021atlite} Technology costs are from NLR's 2022 Annual Technology Baseline\cite{nrel_atb2022} and existing generator data from EIA-860, both accessed through the Public Utility Data Liberation (PUDL) project,\cite{pudl_data} and fuel prices are from the EIA API.\cite{eia_texas_fuel}
    \item Weather data for both systems were generated using Copernicus Climate Change Service information 2026. Neither the European Commission nor ECMWF is responsible for any use that may be made of the Copernicus information or data it contains.
    \item Any additional information required to reanalyze the data reported in this paper is available from the lead contact upon request.
\end{itemize}

\section*{ACKNOWLEDGMENTS}
We would like to thank Samantha Coday for discussions on reactive power modeling, including the suggestion of a dispatch-coupled formulation. We thank Dharik Mallapragada for contributions to the early stages of this project, and Yu Weng, who we collaborated with on a related submission,\cite{weng2026modeling} for fruitful discussions. Shiva Madadkhani thanks Svetlana Ikonnikova for her support in facilitating the research visit to MIT and the resulting collaboration. This publication was based upon work supported by the U.S. Department of Energy’s Office of Energy Efficiency and Renewable Energy (EERE) under the Hydrogen Fuel Cell Technology Office, Award Number DE-EE0010724, and was additionally supported by funding from the MIT Energy Initiative (MITEI).

\section*{AUTHOR CONTRIBUTIONS}
Conceptualization, L.B., R.M., and P.L.D.; methodology, L.B., R.M., P.L.D.; code implementation, L.B.; data preparation, L.B. and S.M; formal analysis, L.B. and S.M.; writing -- original draft, L.B. and S.M.; writing -- review \& editing, L.B., S.M., R.M., and P.L.D.; supervision, R.M. and P.L.D.

\section*{DECLARATION OF INTERESTS}
The authors declare no competing interests.

\section*{DECLARATION OF GENERATIVE AI AND AI-ASSISTED TECHNOLOGIES}
During the preparation of this work, the authors used generative AI tools in order to assist with writing code and debugging data-processing and plotting code. AI was also used to edit excerpts of the manuscript for clarity and grammar. After using this tool, the authors reviewed and edited the content as needed and take full responsibility for the content of the publication.

\section*{SUPPLEMENTAL INFORMATION INDEX}
Document S1: Supplemental Notes S1--S7, Figures S1--S8, Tables S1--S6.

\newpage

\putbib
\end{bibunit}

\newpage
\setcounter{table}{0}
\setcounter{figure}{0}
\renewcommand{\thetable}{S\arabic{table}}
\renewcommand{\thefigure}{S\arabic{figure}}
\renewcommand{\theHtable}{S\arabic{table}}
\renewcommand{\theHfigure}{S\arabic{figure}}

\section*{SUPPLEMENTAL INFORMATION}
\begin{bibunit}

\sisection{Choosing representative periods}\label{si:tdr}

\begin{figure}[htbp]
    \centering
    \includegraphics[width=\linewidth]{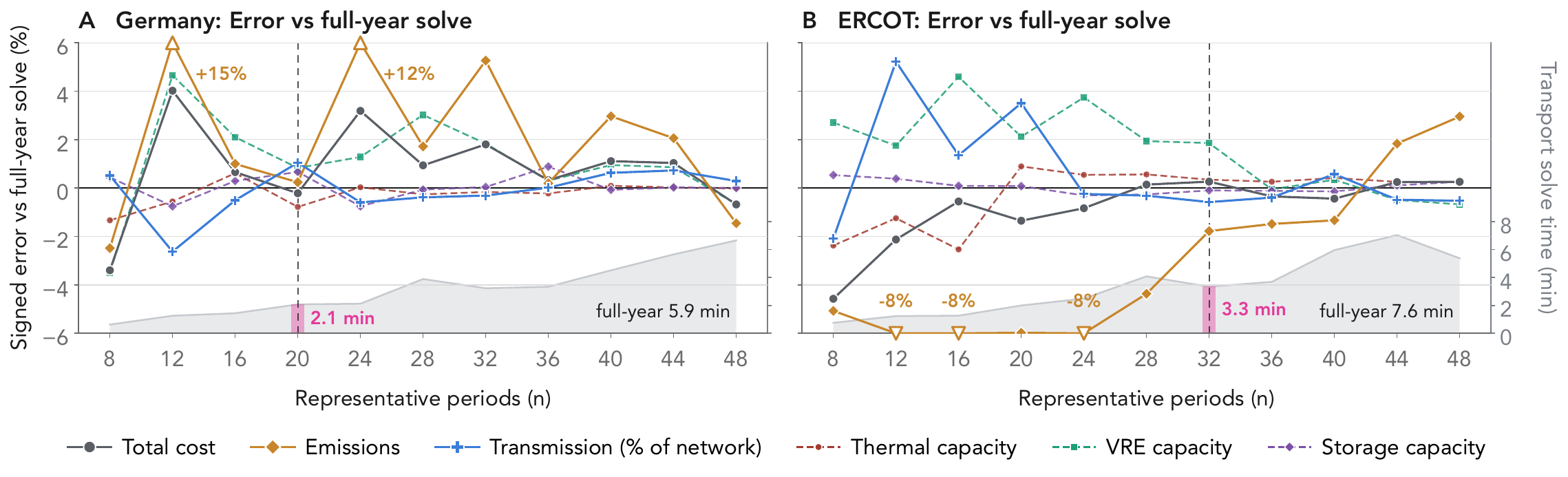}
    \caption{\textbf{Time domain reduction periods vs. solve times.}
    Deviation from the full-year transport solution in system cost, CO$_2$ emissions, delivered transmission capacity, and per-technology capacity, across representative period counts from 8--48 for both systems. Highlighted segment marks the counts used in our results: 20 for Germany and 32 for ERCOT.}
    \label{fig:TDR}
\end{figure}

Solving a full-year CEM involves coupling investment to operations across all 8{,}760 hours of the year. This is easily tractable for the transport formulation but not, within practical compute budgets, for ACOPF, which already requires around ten hours at the representative-period count we use for ERCOT (Table~\ref{tab:solve_times}). To counter intractability in the temporal domain, we use time domain reduction (TDR) in all runs. As mentioned previously, this involves clustering the annual time series for, \textit{e.g.}, demand, fuel prices, and generator variability into a smaller set of representative periods whose weighted sum reproduces a full year of operation. In this section we justify the period counts used in our experiments. 

Both systems use 168-hour (one-week) representative periods. We solve Germany with 20 periods and ERCOT with 32 periods. These period counts sample 38\% and 61\% of the modeled year, respectively. We claim that these choices adequately sample the behavior of the full-year solve. To pick a period count, we solved the lossy transport model with network expansion at period counts of 8--48 in steps of 4, plus a full year, for both systems (Fig.~\ref{fig:TDR}). At the counts we use, the transport solution sits within 2\% of its own full-year counterpart on total system cost, CO$_2$ emissions, delivered transmission capacity (as a percentage of the resulting network), and per-technology capacities (as a percentage of total installed capacity for VRE, thermal, and storage). We selected a lower period count for Germany than for ERCOT because its errors fall inside the same band at 20 periods and show no systematic improvement beyond it. 

Because our main conclusions rest on differences between ACOPF, DCOPF, and transport formulations rather than on any single solution, what matters is whether those differences are stable across representative period counts that sample a sufficient portion of the year. Table~\ref{tabS:n20n32} solves all three formulations for ERCOT at both 20 and 32 periods under a 5\,GW zonal thermal capacity cap. Absolute levels shift between the 20 and 32 period counts: ACOPF's installed storage falls from 4.06 to 2.56\,GW. The differences between formulations do not. DCOPF and transport underbuild storage by 53\% and 55\% at $n=20$ and by 38\% at $n=32$; they build about 2--5\% more transmission than ACOPF at both counts; and agree with ACOPF on total system cost to within 0.25\%. The choice of period count shifts the absolute results of all three formulations, but not the differences between them, which is the basis of our conclusions.

\begin{table}[htbp]
\centering
\small
\caption{\textbf{ERCOT main results metrics at 20 and 32 representative periods, all three
formulations.} Percentages in the lower block are deviations from ACOPF at the same period
count.}
\label{tabS:n20n32}
\begin{tabular}{|l | r r r | r r r|}
\hline
 & \multicolumn{3}{c|}{\textbf{$\mathbf{n =20}$}} & \multicolumn{3}{c|}{\textbf{$\mathbf{n =32}$}} \\
 & Transport & DCOPF & ACOPF & Transport & DCOPF & ACOPF \\ [1ex]
\hline
\multicolumn{7}{|l|}{\textbf{Levels}} \\
\hline
Total cost, bn \$        & 18.97 & 19.04 & 19.01 & 19.25 & 19.32 & 19.27 \\
Total capacity, GW       & 142.0 & 142.2 & 144.7 & 140.6 & 141.0 & 142.7 \\
New transmission, GVA    & 13.69 & 13.35 & 13.14 & 11.32 & 11.43 & 10.93 \\
Storage power, GW        &  1.84 &  1.89 &  4.06 &  1.60 &  1.58 &  2.56 \\
\hline
\multicolumn{7}{|l|}{\textbf{Deviation from ACOPF at the same $\mathbf{n}$}} \\
\hline
Total cost               & $-$0.25\% & $+$0.13\% & -- & $-$0.09\% & $+$0.25\% & -- \\
Total capacity           & $-$1.9\%  & $-$1.7\%  & -- & $-$1.5\%  & $-$1.2\%  & -- \\
New transmission         & $+$4.2\% & $+$1.6\% & -- & $+$3.6\% & $+$4.6\% & -- \\
Storage power            & $-$54.7\% & $-$53.4\% & -- & $-$37.6\% & $-$38.4\% & -- \\
\hline
\end{tabular}
\end{table}

\sisection{Zonal thermal capacity caps in ERCOT}\label{si:caps}

\begin{figure}[htbp]
    \centering
    \includegraphics[width=\linewidth]{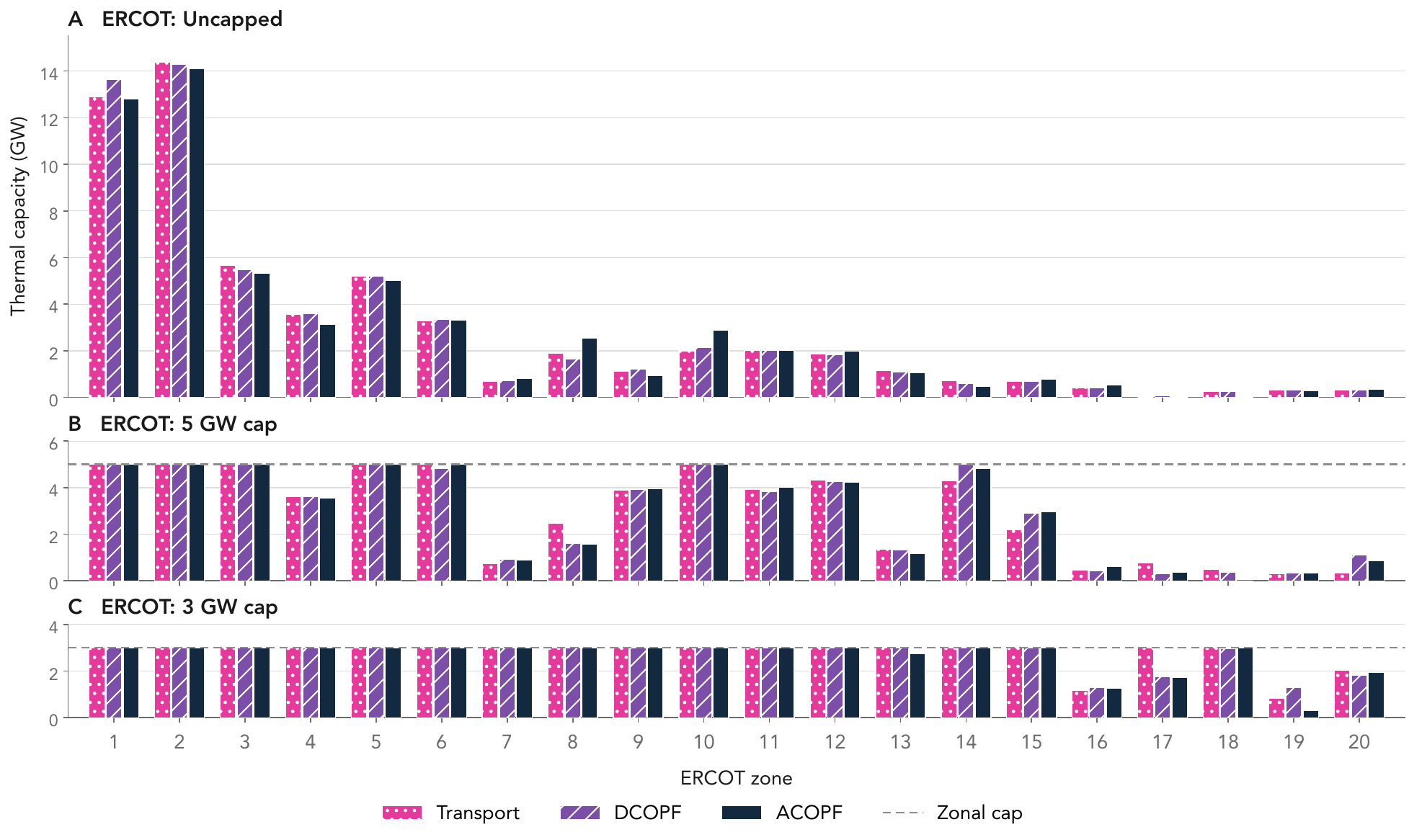}
    \caption{\textbf{ERCOT: Installed thermal capacity under per-zone thermal caps.}
    (A) Uncapped, (B) 5\,GW cap (used for main-text ERCOT results), and (C) 3\,GW cap. Thermal capacity is the combined per-zone sum of CCGT, OCGT and coal (no coal was built under any cap). Dashed line marks the cap. Runs use 20 representative periods with network expansion enabled.}
    \label{fig:thermal_caps}
\end{figure}

For our main-text ERCOT results, we apply a 5\,GW thermal (CCGT and OCGT) per-zone new capacity cap. When we do not apply this cap, the model builds approximately 14\,GW of thermal capacity in each of zones 1 and 2 (Houston and Dallas), strongly co-locating generation with demand rather than distributing it across the network. Peak demand is concentrated in these two zones, which together carry 43\% of ERCOT's annual demand, and incident transmission capacity does not meet peak demand in 8 of 20 zones across the system. With unlimited thermal capacity additions and low fuel prices compared to Germany (Table~\ref{tabS:inputs}), the model chooses to build cheap CCGT at its load centers. Neither ERCOT's planning practice nor its development pipeline behaves this way in reality. ERCOT sites new generation at a county level considering population density, environmental constraints, water availability, and EPA non-attainment areas,\cite{ercot2014ltta} and its scenarios from 2014 place 20--33\,GW of new gas across many dispersed counties rather than in metropolitan load zones. The current installation pipeline has 129 planned gas projects totaling 73\,GW, with its largest (1--2\,GW) plants all in sparsely populated areas.\cite{cleanview_ercot_gas} Air quality is one of the most significant reasons behind avoiding busy metropolitan areas for gas generation siting. The Dallas, Houston, and San Antonio areas all exceed federal ozone limits,\cite{epa_greenbook_tx} and Texas law caps how much NO$_\text{x}$ a gas turbine may emit in those areas.\cite{tceq_nox_egu} We impose a uniform per-zone thermal cap in order to approximate these limiting factors.

Figure~\ref{fig:thermal_caps} compares the uncapped case against both 5\,GW and 3\,GW caps for all three formulations. The 5\,GW cap binds in the 6 largest thermal zones (including Dallas, Houston, and San Antonio) and redistributes that installed thermal capacity among remaining zones. The 3\,GW cap binds in 15 of 20 zones, including several zones which represent dispersed non-urban areas without strict air quality constraints. We did not apply a cap for Germany because its demand is much more distributed (the two largest zones carry 17\% compared to ERCOT's 43\%) and high fuel prices already encourage the system to leverage its low-cost renewables and transmission instead of gas.

\begin{figure}[htbp]
    \centering
    \includegraphics[width=\linewidth]{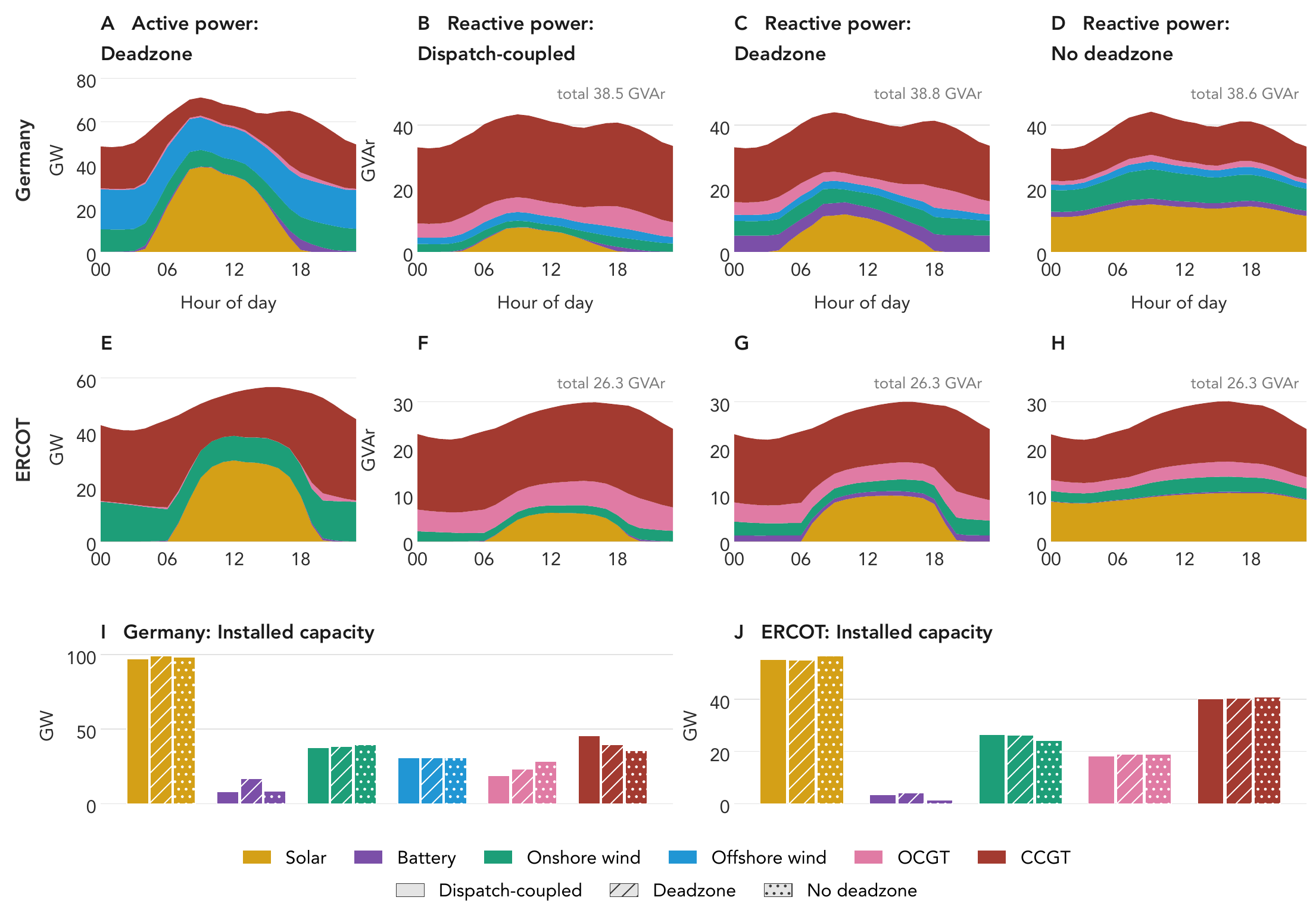}
    \caption{\textbf{Reactive power dispatch and installed capacity under three reactive capability limits (ACOPF).}
    (A--D) Germany; (E--H) ERCOT under the 5\,GW per-zone thermal cap. (A, E) Active power dispatch by technology under the baseline dead-zone limit. (B--D, F--H) Reactive power dispatch ($|Q|$) by technology under the dispatch-coupled, dead-zone (baseline) and no-dead-zone limits. Columns are ordered by how far reactive output is decoupled from active output. All hourly panels show the weighted mean across representative periods for each hour of the day. The label on each reactive panel gives the weighted-mean daily system $|Q|$, which varies by less than 1\% between limits in both systems. (I, J) Installed capacity by technology under each limit. All runs use 20 representative periods with network expansion.}

    \label{fig:q_envelope}
\end{figure}

Another observation is that under the tightest, 3\,GW cap, the ERCOT system begins to more closely resemble Germany (Table~\ref{tabS:caps_vre}). As the cap tightens, the renewable share rises, from 47--48\% under the 5\,GW cap to 52--53\%, and thermal capacity is pushed further from the demand centers it would otherwise serve. Both affect the formulations' disagreement, through different mechanisms. The approximations' siting disagreement with ACOPF rises to levels similar to those in Germany, even though ERCOT's renewable share remains well below Germany's roughly 70\%. This suggests that siting disagreement is driven mainly by the separation of supply from demand, where the physics of how power is routed through the network shapes generation siting and transmission buildout. The storage gap, however, is driven by the renewable share. ACOPF builds 2.7--3.6 times as much battery capacity as the approximations under the 3\,GW cap, and none in any formulation without a cap. Under dead-zone-enabled reactive power limits, inverter-based resources cannot supply reactive power at low output, so a higher renewable share, combined with less synchronous generation near load, leaves ACOPF to meet reactive demand with short-duration batteries (Fig.~\ref{fig:q_envelope}).

\begin{table}[htbp]
\centering
\small
\caption{\textbf{Renewable share, siting disagreement, and storage under different ERCOT thermal caps, compared with Germany.} Renewable share is wind and solar generation as a share of total generation, given as the range across formulations. Siting disagreement is the sum of absolute zone-by-technology capacity differences from ACOPF, as a share of ACOPF's total capacity. All cases use 20 representative periods with network expansion.}
\label{tabS:caps_vre}
\begin{tabular}{|l | c | c c | c c c|}
\hline
 & \textbf{Renewable} & \multicolumn{2}{c|}{\textbf{Siting disagreement}} & \multicolumn{3}{c|}{\textbf{Battery power, GW}} \\
\textbf{Case} & \textbf{share} & Transport & DCOPF & Transport & DCOPF & ACOPF \\ [1ex]
\hline
Germany & 68--69\% & 35.9\% & 31.8\% & 7.3 & 8.1 & 16.6 \\
ERCOT, uncapped & 45--46\% & 18.1\% & 13.7\% & 0.0 & 0.0 & 0.0 \\
ERCOT, 5\,GW cap & 47--48\% & 17.8\% & 10.0\% & 1.8 & 1.9 & 4.1 \\
ERCOT, 3\,GW cap & 52--53\% & 35.0\% & 28.1\% & 3.8 & 5.2 & 13.8 \\
\hline
\end{tabular}
\end{table}

\sisection{Sensitivity to network expansion}\label{si:nonx}

\begin{figure}[htbp]
    \centering
    \includegraphics[width=\linewidth]{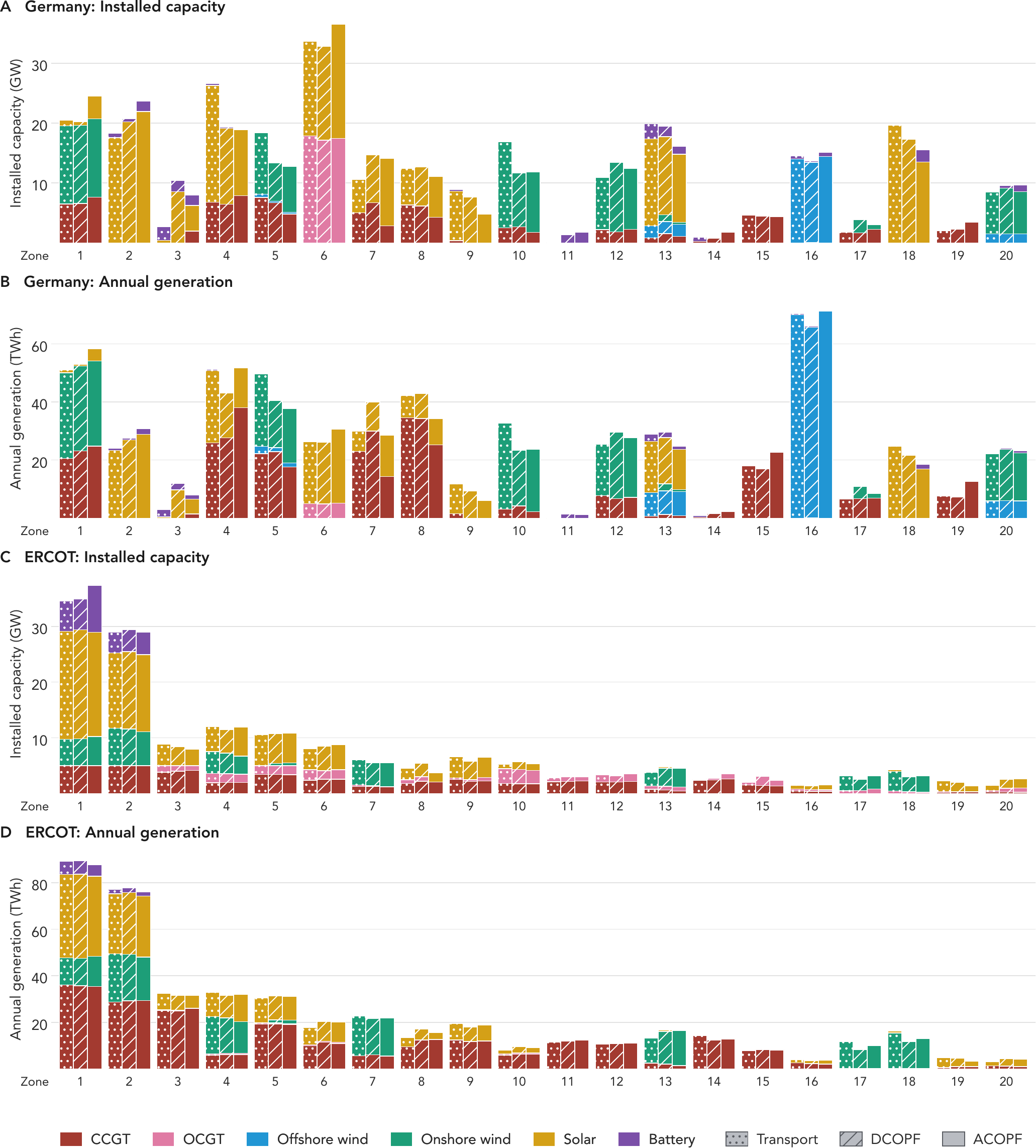}
    \caption{\textbf{Zonal capacity and generation without transmission expansion.}
    (A, B) Germany; (C, D) ERCOT under the 5\,GW per-zone thermal cap. The network is fixed at its existing capacity, and both systems use 20 representative periods. Bars are stacked by technology and grouped by formulation within each zone: ACOPF solid, DCOPF diagonal hatch, transport dotted. Battery generation is gross discharge. Aggregate results are in Table~\ref{tabS:nonx}.}
    \label{figS:noNX_zonal}
\end{figure}

All results in the main text allow the model to build new transmission. In this section, we test how much the formulations' observed differences from ACOPF depend on that degree of freedom. We repeat the three-formulation comparison with the transmission lines fixed at their existing capacities (Fig.~\ref{figS:noNX_zonal}, Table~\ref{tabS:nonx}). We note that without network expansion, the DCOPF formulation becomes linear and solves significantly more quickly than expansion-enabled DC, which is a nonlinear program. For this sensitivity analysis, we use 20-period TDR for both Germany and ERCOT. Because our ERCOT comparisons are all against 20-period runs with network expansion, the numbers in this section should not be differenced against main-text Table~\ref{tab:headline}, whose ERCOT results use 32 periods (see Table~\ref{tabS:n20n32}, which compares several metrics from the two period counts for ERCOT).

\textbf{System-wide agreement with zonal disagreement persists.} Without network expansion, Germany's median zone differs from ACOPF by 25.7\% installed capacity under transport and its median zone's technology mix by 41.4\%. The total system-wide differences in capacity installation in Germany are far smaller at just 1.0\% for transport. In both systems, transport and DC approximations continue to build approximately the right amount of capacity for the system, but the wrong quantity in most individual zones (see Table~\ref{tabS:nonx}). Fixing the network narrows the capacity siting gap in Germany without closing it: misallocation summed across zone--technology pairs falls from 35.9 to 31.3\% under transport when we fix the network and from 31.8 to 22.6\% under DCOPF. In ERCOT, this gap barely moves (17.8 to 14.0\%, 10.0 to 10.1\%). In other words, the zonal capacity and generation disagreement from ACOPF that we see in our main analysis is not due to the formulations building different networks.  

\textbf{The two systems substitute for transmission differently, but ACOPF installs more, shorter duration batteries in both.} When ERCOT cannot build new transmission, it turns to storage in all three formulations, increasing installed battery power capacity by 3--5$\times$ in place of the 13--14\,GVA of new transmission it otherwise builds. Germany relocates its generation instead. With network expansion, ACOPF concentrates 27.1\,GW of offshore wind in zone 16, and all three lines leaving that zone are maximally reinforced. Without network expansion, that capacity falls to 14.4\,GW. The difference is made up by onshore wind, solar, and CCGT built closer to its load centers, while storage actually \textit{falls} rather than rising. Offshore wind is Germany's most productive resource in these models, at a mean capacity factor of 0.49 against 0.25 for onshore wind and 0.13 for solar, but without reinforcing the lines out of these offshore-rich zones, the model cannot deliver it and instead builds closer to demand. ERCOT's wind is also remote, but solar is strong in most zones (Fig.~\ref{fig:vre_maps}), so storage can replace transmission capability more easily. In both systems, ACOPF builds more battery power capacity than the approximations, and at shorter durations. Even without network expansion enabled, ACOPF still has to serve reactive power demand that the other formulations cannot see, so our previous analysis of storage carries over to this setting. 

\begin{table}[htbp]
\centering
\caption{\textbf{Effect of removing transmission expansion, by formulation.} Cost, capacity, zonal disagreement and solve time.}
\label{tabS:nonx}

\noindent\begin{minipage}{\linewidth}
\centering
\scriptsize
\setlength{\tabcolsep}{4pt}
\setlength{\parskip}{0pt}
\textbf{Germany, 20 periods}

\medskip
\begin{tabular}{|l | r r | r r | r r|}
\hline
 & \multicolumn{2}{c|}{\textbf{Transport}} & \multicolumn{2}{c|}{\textbf{DCOPF}} & \multicolumn{2}{c|}{\textbf{ACOPF}} \\
 & \multicolumn{1}{c}{\textbf{NX}} & \multicolumn{1}{c|}{\textbf{noNX}} & \multicolumn{1}{c}{\textbf{NX}} & \multicolumn{1}{c|}{\textbf{noNX}} & \multicolumn{1}{c}{\textbf{NX}} & \multicolumn{1}{c|}{\textbf{noNX}} \\ [1ex]
\hline
\multicolumn{7}{|l|}{\textbf{Cost (billion \EUR{}) and emissions}} \\ [1ex]
\hline
\textbf{Total} & 38.67 & 39.59 & 39.13 & 40.04 & 38.66 & 39.44 \\
Emissions (Mt CO$_2$) & 60.3 & 65.5 & 62.5 & 69.6 & 59.6 & 67.3 \\
\hline\hline
\multicolumn{7}{|l|}{\textbf{Capacity (GW)}} \\ [1ex]
\hline
Thermal (CCGT + OCGT) & 64.7 & 64.6 & 64.3 & 64.9 & 62.6 & 63.6 \\
Offshore wind & 29.3 & 18.1 & 28.7 & 17.1 & 30.9 & 18.2 \\
Onshore wind & 43.2 & 53.4 & 41.3 & 51.1 & 38.4 & 49.2 \\
Solar & 97.8 & 108.3 & 101.9 & 110.1 & 99.0 & 107.9 \\
Storage (battery) & 7.3 & 7.4 & 8.1 & 6.6 & 16.6 & 10.4 \\
\quad\textit{duration, h} & 2.91 & 2.83 & 2.82 & 2.82 & 1.60 & 2.18 \\
\hline
Total capacity & 242.3 & 251.8 & 244.3 & 249.8 & 247.4 & 249.4 \\
\hline\hline
\multicolumn{7}{|l|}{\textbf{Zonal disagreement with ACOPF} (\% of ACOPF capacity)} \\ [1ex]
\hline
Misallocation, all zone-tech pairs & 35.9 & 31.3 & 31.8 & 22.6 & -- & -- \\
Median zone, total capacity & 37.5 & 25.7 & 12.0 & 11.7 & -- & -- \\
Median zone, technology mix & 52.8 & 41.4 & 27.6 & 20.5 & -- & -- \\
\hline\hline
\multicolumn{7}{|l|}{\textbf{Solve time}} \\ [1ex]
\hline
Wall-clock (s) & 126 & 355 & 10,145 & 548 & 18,040 & 16,806 \\
Speed-up vs.\ ACOPF & 144$\times$ & 47$\times$ & 2$\times$ & 31$\times$ & 1.0$\times$ & 1.0$\times$ \\
\hline
\end{tabular}
\end{minipage}

\nopagebreak
\vspace{1.2em}
\nopagebreak

\noindent\begin{minipage}{\linewidth}
\centering
\scriptsize
\setlength{\tabcolsep}{4pt}
\setlength{\parskip}{0pt}
\textbf{ERCOT, 5\,GW per-zone thermal cap, 20 periods}

\medskip
\begin{tabular}{|l | r r | r r | r r|}
\hline
 & \multicolumn{2}{c|}{\textbf{Transport}} & \multicolumn{2}{c|}{\textbf{DCOPF}} & \multicolumn{2}{c|}{\textbf{ACOPF}} \\
 & \multicolumn{1}{c}{\textbf{NX}} & \multicolumn{1}{c|}{\textbf{noNX}} & \multicolumn{1}{c}{\textbf{NX}} & \multicolumn{1}{c|}{\textbf{noNX}} & \multicolumn{1}{c}{\textbf{NX}} & \multicolumn{1}{c|}{\textbf{noNX}} \\ [1ex]
\hline
\multicolumn{7}{|l|}{\textbf{Cost (billion \$) and emissions}} \\ [1ex]
\hline
\textbf{Total} & 18.97 & 19.93 & 19.04 & 20.02 & 19.01 & 19.91 \\
Emissions (Mt CO$_2$) & 104.5 & 97.3 & 105.9 & 99.0 & 102.8 & 98.8 \\
\hline\hline
\multicolumn{7}{|l|}{\textbf{Capacity (GW)}} \\ [1ex]
\hline
Thermal (CCGT + OCGT) & 59.3 & 53.6 & 59.8 & 54.9 & 59.4 & 54.5 \\
Wind (onshore) & 24.8 & 28.7 & 24.0 & 27.4 & 26.2 & 28.1 \\
Solar & 56.1 & 61.1 & 56.5 & 62.2 & 55.0 & 60.6 \\
Storage (battery) & 1.8 & 9.2 & 1.9 & 9.5 & 4.1 & 12.5 \\
\quad\textit{duration, h} & 1.68 & 4.36 & 1.63 & 4.26 & 1.03 & 2.80 \\
\hline
Total capacity & 142.0 & 152.6 & 142.2 & 154.0 & 144.7 & 155.6 \\
\hline\hline
\multicolumn{7}{|l|}{\textbf{Zonal disagreement with ACOPF} (\% of ACOPF capacity)} \\ [1ex]
\hline
Misallocation, all zone-tech pairs & 17.8 & 14.0 & 10.0 & 10.1 & -- & -- \\
Median zone, total capacity & 10.0 & 7.9 & 3.7 & 6.7 & -- & -- \\
Median zone, technology mix & 17.2 & 15.9 & 9.5 & 10.9 & -- & -- \\
\hline\hline
\multicolumn{7}{|l|}{\textbf{Solve time}} \\ [1ex]
\hline
Wall-clock (s) & 121 & 94 & 8,590 & 324 & 21,666 & 12,599 \\
Speed-up vs.\ ACOPF & 179$\times$ & 134$\times$ & 3$\times$ & 39$\times$ & 1.0$\times$ & 1.0$\times$ \\
\hline
\end{tabular}
\end{minipage}
\end{table}

\sisection{Sensitivity to reactive power capability limits}\label{si:qlimits}

We compare three reactive generation limits (Fig.~\ref{fig:q_envelope}). All three limit a generator's reactive power capability with a minimum rated power factor $\mathrm{PF}$, the
ratio of active to apparent power the generator must sustain. We constrain reactive output up to
$\mathcal{Q}= \tan(\arccos(\mathrm{PF}))$ times a given active quantity to which the limit is anchored. We use $\mathrm{PF} = 0.85$ for thermal resources and the tighter $\mathrm{PF} = 0.95$ for
inverter-based resources (IBRs), giving a reactive factor $\mathcal{Q}$ of 0.62 and 0.33 respectively (see \hyperref[methods:qlimits]{Methods, \emph{Reactive power capacity limits}}). The three models differ in what we apply this factor to:

\begin{itemize}
    \item \textbf{No dead-zone}: Reactive generation is bounded symmetrically by $\pm\mathcal{Q}$ times installed capacity, per resource type and zone.
    \item \textbf{Dead-zone} (baseline): As above for thermal devices, but IBRs cannot generate reactive power below 5\% capacity factor, ramping up linearly to the full limit at 20\%.
    \item \textbf{Dispatch-coupled}: As above for thermal; IBRs are bounded by $\mathcal{Q}$ times active dispatch, so idle equipment provides no reactive support.
\end{itemize}

We study these limits because grid requirements vary: the codes in Germany require a dead-zone,\cite{vde4120} while ERCOT expects support across the full range from IBRs interconnecting after 2024.\cite{ercot2026reactive} The dispatch-coupled case is included from a conservative, device-level perspective, but does not follow modern requirements.\cite{vde4120, ieee2800} We apply the dead-zone to both systems in our main results and study the other limits as a sensitivity. Reactive demand is a fixed input, so these limits vary only in how the network meets it. 

Without a dead-zone, solar's reactive output roughly triples in Germany and doubles in ERCOT. Most of that displaced reactive support comes from thermal generators, but only because they are the largest reactive source. Storage gives up the largest fraction of reactive support, falling 62\% in Germany and 81\% in ERCOT, compared to CCGT's drop of 36\% and 19\%. When solar can provide reactive support at any irradiance, installing batteries to sit idle for nighttime reactive capability stops being worthwhile. 

ERCOT's storage capacity falls from 4.1\,GW with a dead-zone to 1.4\,GW without and 3.3\,GW under dispatch-coupled limits. Germany's roughly halves under both alternatives, going from 16.6\,GW to 8.5 and 8.0\,GW. Storage \textit{energy} barely moves, so the system stops building short-duration capacity. Mean duration is 1.0, 1.3 and 2.0\,h in ERCOT and 1.6, 2.8 and 2.5\,h in Germany under the dead-zone, without it, and dispatch-coupled, respectively---shortest under the dead-zone in both. 

Thermal reactive capability is anchored on installed capacity in all cases, so changes observed in thermal resources are due to differences in how much reactive support IBRs can take on. Total gas capacity is nearly constant in both systems. In Germany, thermal plants supply 35\% of system reactive output without the dead-zone, 58\% with it, and 78\% under the dispatch-coupled limit. Under the dispatch-coupled limit, more CCGT is built to meet reactive demand when IBRs are idle, raising emissions overall. ERCOT's gas fleet stays nearly unchanged, with its response likely dampened by the 5\,GW zonal thermal cap binding in 6 high-demand zones.

Changes in the system's total installed capacity are mostly from storage differences. Germany's installed capacity drops from 247\,GW under the dead-zone to 240\,GW without it and 238\,GW under the dispatch-coupled limit. In fact, Germany's system-wide capacity is more sensitive to reactive limits than to the power flow formulation, with DCOPF and transport totals 1.2\% and 2.1\% below the dead-zone ACOPF baseline against 2.8\% and 3.9\% for the two alternative limits. At a zonal level, the share of ACOPF capacity the approximations allocate differently stays between 20\% and 40\% in Germany and between 10\% and 22\% in ERCOT under all three limits, while system-wide totals agree to within 3\%. Choice of reactive generation limit changes how much storage ACOPF builds relative to the approximations, but they still build the wrong technologies in the wrong places.

\sisection{Sensitivity to reactive supply and demand power factors}\label{si:pfsweep}

\begin{figure}[htbp]
    \centering
    \includegraphics[width=\linewidth]{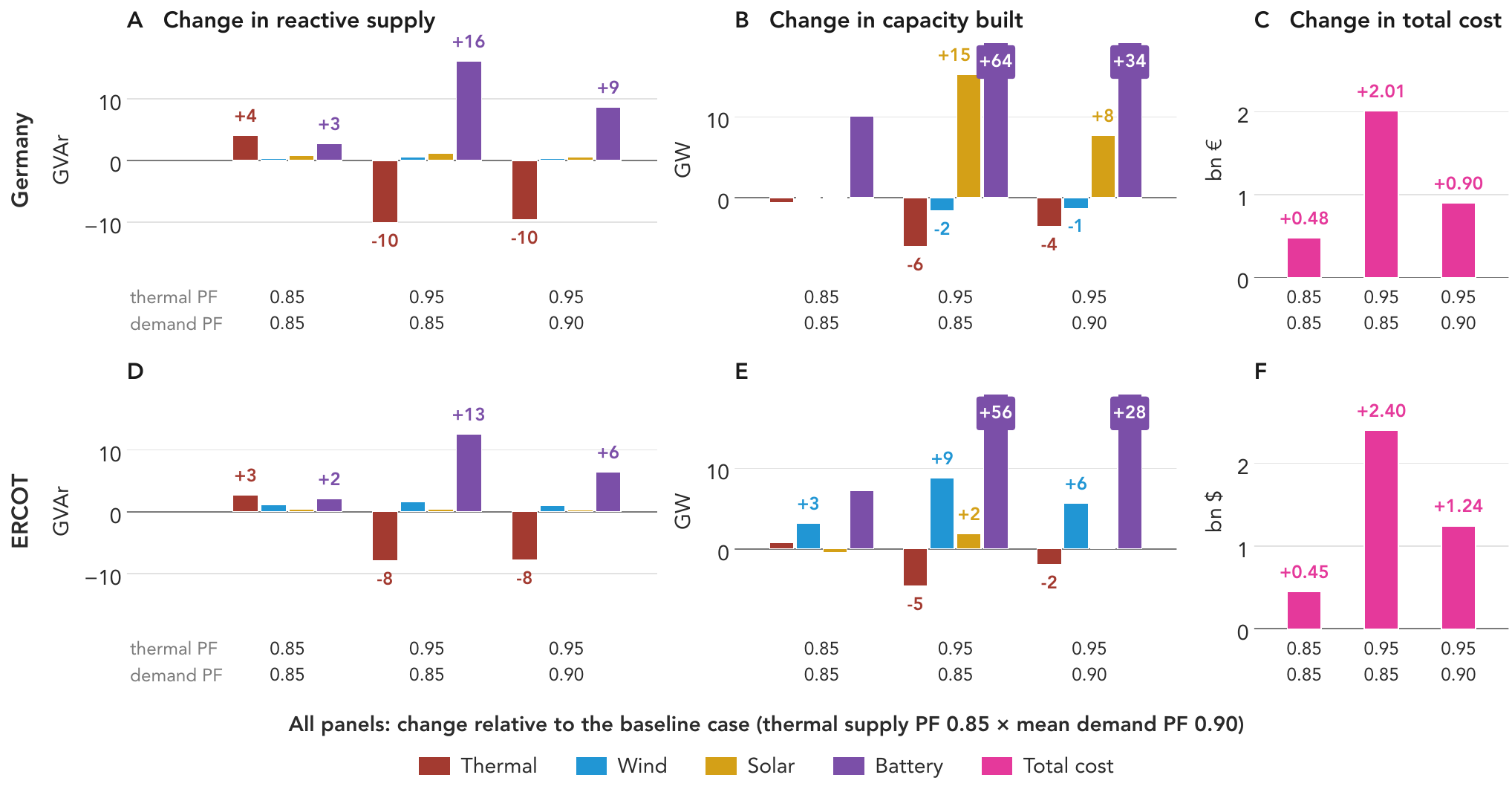}
    \caption{\textbf{Sensitivity analysis for reactive supply and demand power factors.} All subfigures show absolute change against the baseline (0.85 thermal limit PF, 0.9 mean demand PF). Reactive generation limits all follow the dead-zone model from our baseline. Germany and ERCOT are both modeled with 20 representative periods.}
    \label{fig:q_sweep}
\end{figure}

The results in the main text all follow the same power factor assumptions: thermal generators at 0.85, IBRs at 0.95, and mean demand at 0.9. We perform a $2\times2$ sweep over thermal power factors of (0.85, 0.95) and mean demand power factors of (0.85, 0.90), keeping the IBRs' power factor fixed, to analyze the system's sensitivity to reactive power-constrained scenarios. The demand power factor sets the reactive requirement, raising the total level by around a fifth in our sensitivity. We tighten thermal power factors because synchronous generators have the greatest reactive capability, so constraining them to the same power factor as IBRs simulates a system in which all reactive power is supplied at the same level. 

For the tightest case (0.95 thermal $\times$ 0.85 demand) in both systems, thermal's share of reactive support drops steeply and storage rises sharply, while VRE barely moves (Fig.~\ref{fig:q_sweep}). Storage capacity investment changes by an order of magnitude higher than any other technology, rising from 16.6 to 80.4\,GW in Germany and from 4.1 to 59.9\,GW in ERCOT with durations dropping to the minimum, a flat 1-hour, in both systems. Storage is overwhelmingly providing reactive support and not energy capacity in this stressed reactive-power case. Under our modeling choices which do not include a full range of network-supporting devices, short-duration storage is the most cost-effective resource for providing reactive support. 

The tightest case is not a configuration we would expect to see in real planning settings. This sensitivity exaggerates reactive requirements to highlight that this is a part of the model DC and transport simply cannot see, and the differences from ACOPF grow stronger in these situations (Table~\ref{tabS:pfdiverge}). However, in high-renewables systems or heavy industrial load pockets, conditions closer to these stressed scenarios can occur and are under examination.\cite{nerc_large_loads_2025}  

\begin{table}[htbp]
\centering
\small
\caption{\textbf{Deviation of DCOPF and transport from ACOPF across the reactive sweep.} Cells are (thermal PF $\times$ mean demand PF). Percentages are deviations from the ACOPF solution in the same cell. Germany and ERCOT are at $n=20$ with network expansion, ERCOT under the 5\,GW per-zone thermal cap. \textbf{Note.} ERCOT's cost of non-served energy falls from \$129.3 M at the baseline  to approximately zero in both 0.95 cells, so part of its cost gradient is slack being priced out rather than investment added; Germany's is approximately zero for all runs.}
\label{tabS:pfdiverge}
\begin{tabular}{|l l | r r r r|}
\hline
 & & \multicolumn{4}{c|}{Thermal PF, Demand PF} \\
&& $(0.85, 0.90)$ & $(0.85, 0.85)$ & $(0.95, 0.90)$ & $(0.95, 0.85)$ \\ [1ex]
\hline
\multicolumn{6}{|l|}{\textbf{Germany}} \\
\hline
 & ACOPF storage, GW    & 16.55 & 26.65 & 50.69 & 80.38 \\
 & Storage, DCOPF       & $-$50.9\% & $-$69.5\% & $-$84.0\% & $-$89.9\% \\
 & Storage, transport   & $-$55.6\% & $-$72.5\% & $-$85.5\% & $-$90.9\% \\
 & Total capacity, DCOPF     & $-$1.3\% & $-$4.9\% & $-$14.1\% & $-$23.4\% \\
 & Total capacity, transport & $-$2.1\% & $-$5.7\% & $-$14.8\% & $-$24.0\% \\
 & Total cost, DCOPF         & $+$1.2\% & $-$0.03\% & $-$1.1\% & $-$3.8\% \\
 & Total cost, transport     & $+$0.0\% & $-$1.2\% & $-$2.3\% & $-$4.9\% \\
\hline
\multicolumn{6}{|l|}{\textbf{ERCOT}} \\
\hline
 & ACOPF storage, GW    & 4.06 & 11.33 & 32.29 & 59.86 \\
 & Storage, DCOPF       & $-$53.4\% & $-$83.3\% & $-$94.1\% & $-$96.8\% \\
 & Storage, transport   & $-$54.7\% & $-$83.8\% & $-$94.3\% & $-$96.9\% \\
 & Total capacity, DCOPF     & $-$1.7\% & $-$8.5\% & $-$19.6\% & $-$31.2\% \\
 & Total capacity, transport & $-$1.9\% & $-$8.6\% & $-$19.7\% & $-$31.3\% \\
 & Total cost, DCOPF         & $+$0.2\% & $-$2.2\% & $-$6.0\% & $-$11.1\% \\
 & Total cost, transport     & $-$0.2\% & $-$2.5\% & $-$6.4\% & $-$11.4\% \\
\hline
\end{tabular}
\end{table}

\sisection{System inputs}\label{si:inputs}

\begin{figure}[H]
    \centering
    \includegraphics[width=.9\linewidth]{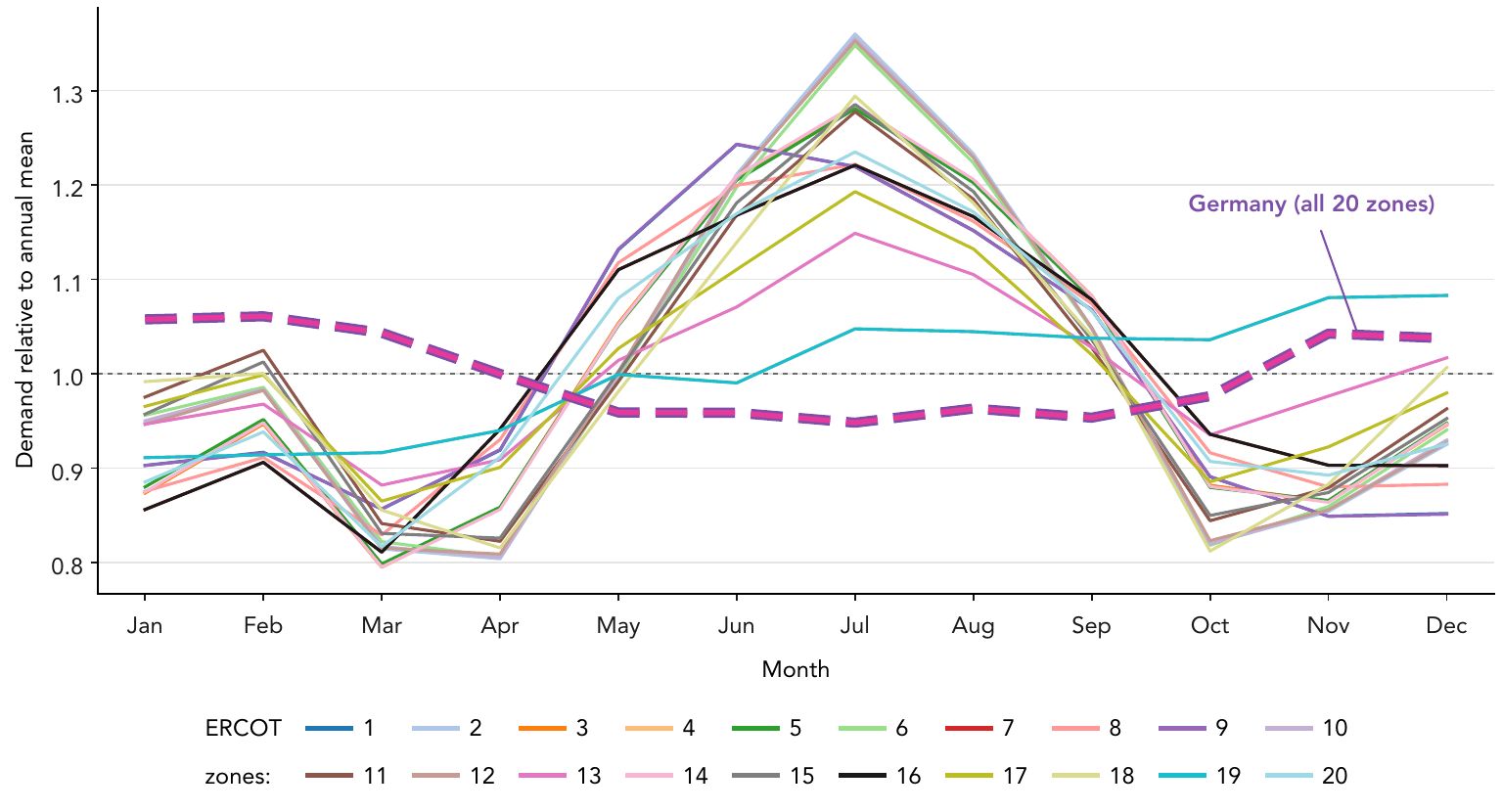}
    \caption{\textbf{Seasonal demand profile across zones}. In ERCOT, electricity demand peaks in warm summer months due to cooling (e.g., air conditioning) needs. In Germany, demand peaks in cooler winter months. All German zones follow the same seasonal profile as a result of the data generation process (see \hyperref[methods:de]{Methods, \emph{Germany}}).}
    \label{figS:demand}
\end{figure}

\begin{figure}[H]
    \centering
    \includegraphics[width=\linewidth]{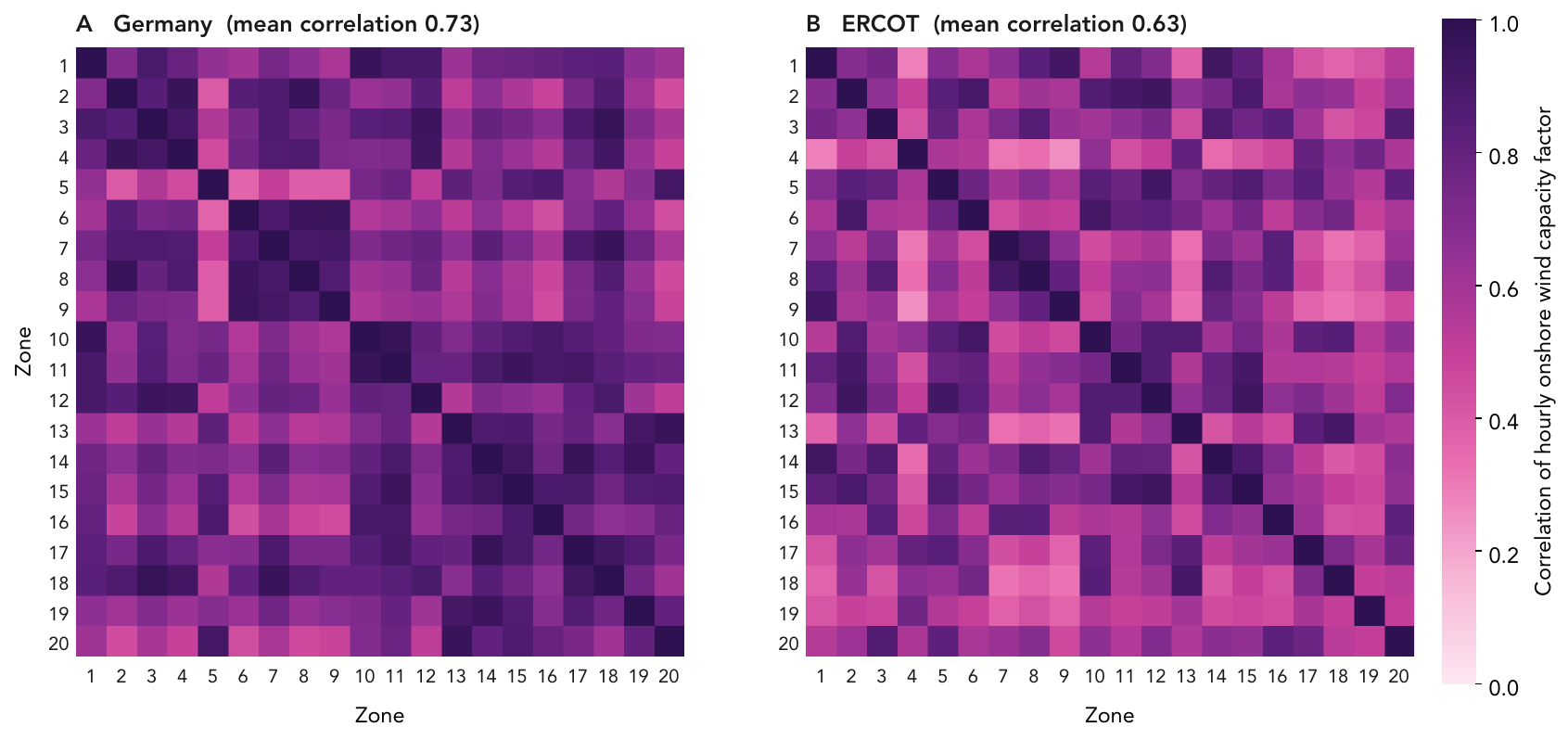}
    \caption{\textbf{Zonal correlation of onshore wind availability.} Correlation between hourly onshore wind capacity factors of each pair of zones annually, for (A) Germany and (B) ERCOT.}
\label{figS:wind_corr}
\end{figure}

This section contains various figures and tables specifying details of the inputs to each system. Table~\ref{tabS:inputs} lists the cost and technical assumptions for each system. Fig.~\ref{figS:demand} shows the seasonal demand profile of each zone, and Fig.~\ref{figS:wind_corr} the correlation of hourly onshore wind capacity factors between zones.

\begin{table}[htbp]
\centering
\caption{\textbf{System input costs and technical assumptions.}}
\label{tabS:inputs}

\begin{center}
\scriptsize
\setlength{\tabcolsep}{4pt}
\begin{tabular}{|l | r | r|}
\hline
\textbf{Parameter} & \textbf{Germany (\EUR{})} & \textbf{Texas / ERCOT (\$)} \\ [1ex]
\hline
\multicolumn{3}{|l|}{\textbf{Annualized investment cost (per MW-yr)}} \\
\hline
CCGT & 95,249 & 59,089 \\
\hline
OCGT & 49,949 & 51,998 \\
\hline
Coal & -- & 175,576 \\
\hline
Onshore wind & 111,607 & 80,167 \\
\hline
Offshore wind & 182,081 & -- \\
\hline
Solar PV & 37,566 & 37,748 \\
\hline
Battery & 11,808 & 15,859 \\
\hline
Battery energy (per MWh-yr) & 32,800 & 18,444 \\
\hline
\multicolumn{3}{|l|}{\textbf{Fixed O\&M (per MW-yr)}} \\
\hline
CCGT & 37,178 & 27,940 \\
\hline
OCGT & 10,358 & 21,430 \\
\hline
Coal & -- & 73,570 \\
\hline
Onshore wind & 16,851 & 42,190 \\
\hline
Offshore wind & 41,809 & -- \\
\hline
Solar PV & 11,890 & 19,952 \\
\hline
Battery & 4,956 & 6,063 \\
\hline
\multicolumn{3}{|l|}{\textbf{Variable O\&M (per MWh) and heat rate}} \\
\hline
CCGT & 5.62 & 1.78 \\
\hline
OCGT & 6.02 & 5.10 \\
\hline
Coal & -- & 7.96 \\
\hline
Onshore wind & 1.81 & 0.00 \\
\hline
Offshore wind & 0.02 & -- \\
\hline
Solar PV & 0.00 & 0.00 \\
\hline
Battery & 0.00 & 0.00 \\
\hline
CCGT heat rate (MMBtu/MWh) & 6.55 & 8.63 \\
\hline
OCGT heat rate (MMBtu/MWh) & 7.44 & 11.84 \\
\hline
Coal heat rate (MMBtu/MWh) & -- & 10.65 \\
\hline
\multicolumn{3}{|l|}{\textbf{Fuel}} \\
\hline
Natural gas (per MMBtu) & 11.98 & 2.45--4.74 \\
\hline
Gas CO$_2$ content (t/MMBtu) & 0.0592 & 0.0529 \\
\hline
Coal (per MMBtu) & -- & 1.82 \\
\hline
Coal CO$_2$ content (t/MMBtu) & -- & 0.0960 \\
\hline
\multicolumn{3}{|l|}{\textbf{Transmission}} \\
\hline
Reinforcement cost (per MW-yr) & 3,712--12,083 & 4,410--23,051 \\
\hline
Line losses (\%) & 3.5 & 2.5 \\
\hline
\multicolumn{3}{|l|}{\textbf{Other}} \\
\hline
Value of lost load (per MWh) & 16,730 & 5,000 \\
\hline
Battery round-trip efficiency & 0.96 & 0.85 \\
\hline
\end{tabular}
\end{center}
\end{table}

\sisection{System outputs}\label{si:outputs}

\begin{figure}[htbp]
    \centering
    \includegraphics[width=\linewidth]{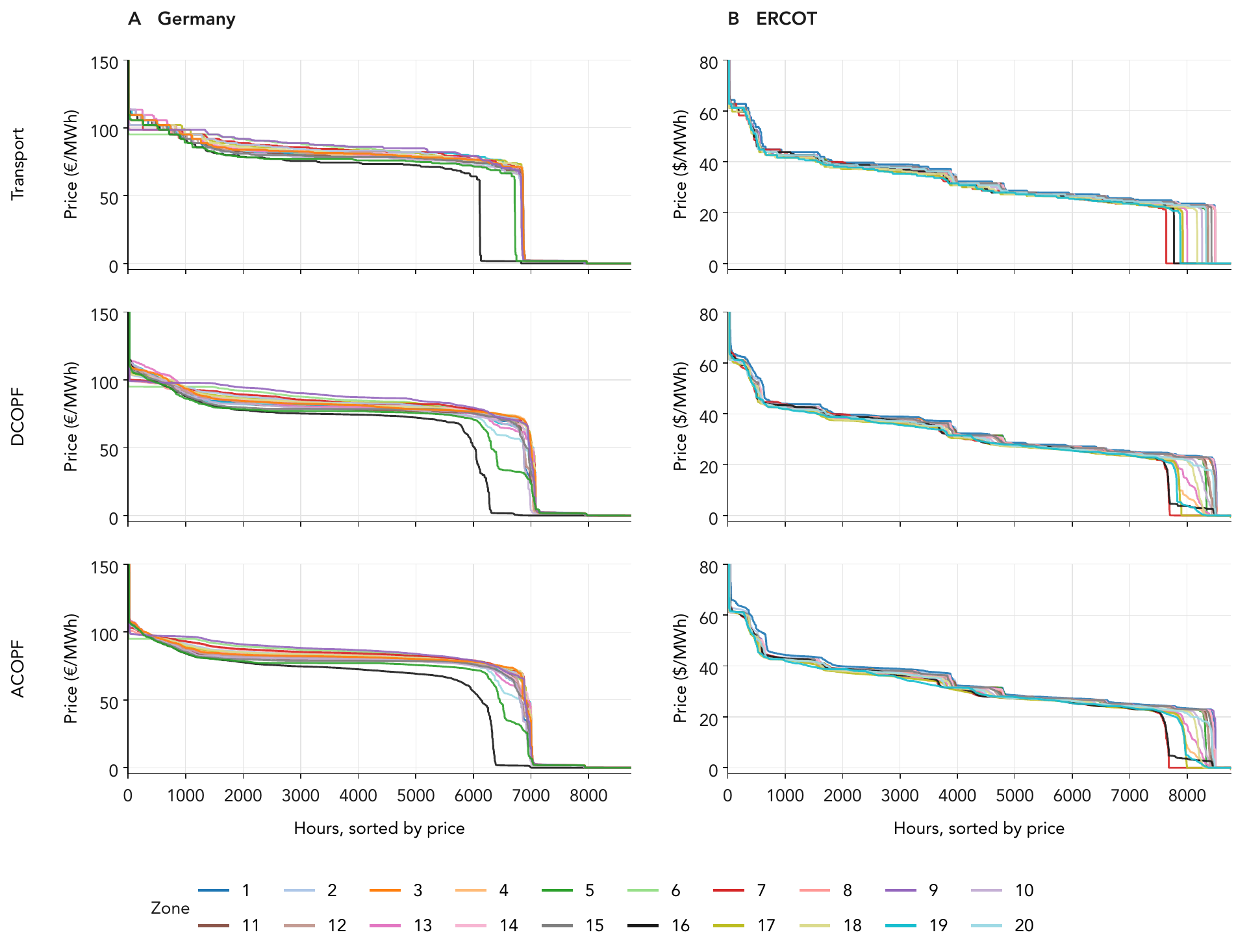}
    \caption{\textbf{Zonal price duration curves by formulation.} (A) Germany and (B) ERCOT under the 5\,GW per-zone thermal cap, for transport, DCOPF and ACOPF (rows). Each curve is one zone's hourly electricity price (the shadow price of its active power balance), sorted from highest to lowest. Each representative hour spans the number of hours it represents under TDR, so the axis covers the full year. Prices are capped at 150\,\EUR{}/MWh (Germany) and 80\,\$/MWh (ERCOT) with 6--25 hours per zone in Germany and 18--48 in ERCOT (under 0.6\% of the year) lying above this range, including a small number of scarcity hours priced near the value of lost load.}
    \label{figS:price_duration}
\end{figure}

Table~\ref{tabS:zonal_balance} gives each zone's annual net export under each formulation, and Fig.~\ref{figS:price_duration} shows price-duration curves of each zone's hourly electricity price, the shadow price of its energy balance.

\begin{table}[htbp]
\centering
\scriptsize
\setlength{\tabcolsep}{4pt}
\caption{\textbf{Annual net export by zone and formulation.} Net export is annual generation plus net storage output minus demand, in TWh; positive values are net exporters and negative values net importers. Demand is an input and identical across formulations. Zone numbers refer to different zones in each system. $^\dagger$Zone is a net exporter under at least one formulation and a net importer under another, counting zones within $\pm$1\,TWh as balanced (as in Fig.~\ref{fig:flows}). Supplied from another zone is $\sum_z|\text{net export}_z|/2$ as a share of total demand; because hours net out over the year, it is a lower bound on hourly transfers.}
\label{tabS:zonal_balance}
\begin{tabular}{|r | r | r r r @{\hspace{4pt}\vrule width 1.2pt\hspace{4pt}} r | r r r|}
\hline
 & \multicolumn{4}{c@{\hspace{4pt}\vrule width 1.2pt\hspace{4pt}}}{\textbf{Germany}} & \multicolumn{4}{c|}{\textbf{ERCOT}} \\
 & & \multicolumn{3}{c@{\hspace{4pt}\vrule width 1.2pt\hspace{4pt}}}{Net export, TWh} & & \multicolumn{3}{c|}{Net export, TWh} \\
\textbf{Zone} & Demand & Transport & DCOPF & ACOPF & Demand & Transport & DCOPF & ACOPF \\
\hline
1 & 50.6 & -2.6 & -0.1 & -5.7 & 97.8 & -28.9 & -27.4 & -30.5 \\
2 & 34.6 & -13.5 & -9.1 & -7.6 & 88.1 & -31.8 & -28.4 & -31.5 \\
3 & 33.8 & -28.4 & -20.7 & -30.0 & 34.4 & -1.5 & -3.3 & -2.7 \\
4 & 30.1 & +23.9 & +14.0 & +23.2 & 33.5 & -5.2 & -6.5 & -3.9 \\
5 & 29.3 & +16.9\rlap{$^\dagger$} & -4.6 & -0.2 & 31.2 & -0.1 & -0.5 & -0.2 \\
6 & 28.9 & +18.2 & +14.1 & +21.8 & 20.9 & -0.6 & +1.9 & +2.5 \\
7 & 28.1 & -6.7 & -4.2 & -19.1 & 18.7 & +9.5 & +7.8 & +9.1 \\
8 & 27.7 & +7.1\rlap{$^\dagger$} & +9.4 & -3.6 & 15.8 & -3.3 & -2.5 & -3.1 \\
9 & 26.5 & -17.9 & -14.2 & -22.6 & 14.4 & +10.8 & +9.6 & +10.1 \\
10 & 24.9 & -13.6 & -5.3 & -3.3 & 13.6 & +4.6 & +5.8 & +7.0 \\
11 & 24.8 & -24.8 & -24.8 & -24.8 & 12.1 & +2.6 & +2.6 & +3.0 \\
12 & 24.5 & +1.9 & +1.4 & +3.7 & 11.6 & +3.7 & +1.5 & +1.6 \\
13 & 22.6 & -4.4 & -2.1 & -2.1 & 10.1 & +0.4 & +4.7 & +2.7 \\
14 & 21.4 & -21.4 & -21.4 & -19.2 & 5.5 & +11.1 & +10.8 & +9.9 \\
15 & 20.9 & -2.5 & -6.6 & -3.2 & 5.0 & +5.6 & +6.3 & +7.9 \\
16 & 19.5 & +106.4 & +103.3 & +112.4 & 4.7 & -1.6 & -1.9 & -1.6 \\
17 & 17.9 & -15.7 & -11.2 & -8.8 & 3.9 & +16.0 & +12.3 & +12.1 \\
18 & 16.3 & -6.1 & -5.6 & -1.8 & 3.1 & +11.3 & +7.1 & +7.6 \\
19 & 13.5 & -5.1 & -4.2 & -0.9 & 2.9 & +1.2 & +1.6 & +0.5 \\
20 & 9.5 & +6.0 & +10.8 & +3.7 & 2.6 & -0.3 & +1.7 & +1.2 \\
\hline
\multicolumn{2}{|l|}{\shortstack[l]{Supplied from\\another zone}} & 34.0\% & 28.4\% & 31.4\% & & 17.5\% & 16.8\% & 17.3\% \\
\hline
\end{tabular}
\end{table}

\renewcommand{\refname}{Supplemental references}
\putbib
\end{bibunit}

\end{document}